\documentclass[
reprint,
nofootinbib,
]{revtex4-2}
\newcommand{\mnras}{Mon. Not. R. Astron. Soc.}
\usepackage{amsmath}
\usepackage{amssymb}
\usepackage{graphicx}
\graphicspath{{figs/}}
\usepackage{xcolor}
\usepackage[colorlinks=true,linkcolor=red,urlcolor=blue,citecolor=blue]{hyperref}
\usepackage{autobreak}
\usepackage{float}
\usepackage{indentfirst}

\begin{document}
\title{Effects of black hole environments on extreme mass-ratio hyperbolic encounters}

\author{Ya-Ze Cheng$^{a}$, Yan Cao$^{b}$, and Yong Tang$^{a,c,d}$}
\affiliation{\begin{footnotesize}
		${}^a$School of Astronomy and Space Sciences, University of Chinese Academy of Sciences (UCAS), Beijing 100049, China\\
        ${}^b$School of Physics, Nanjing University, Nanjing 210093, China\\
		${}^c$School of Fundamental Physics and Mathematical Sciences, \\
		Hangzhou Institute for Advanced Study, UCAS, Hangzhou 310024, China \\
		${}^d$International Center for Theoretical Physics Asia-Pacific, Beijing/Hangzhou, China
\end{footnotesize}}

\date{\today}

\begin{abstract}
Extreme mass-ratio hyperbolic encounters (EMRHEs) around the supermassive black holes will be observable at the future gravitational-wave (GW) detectors in space, such as LISA and Taiji. Here we consider such EMRHEs in the presence of surrounding matter distribution including baryonic accretion disk and dark matter (DM) spike, and estimate their effects on the orbital evolution and GW waveforms. We find that large possible impacts come from the gravitational potential of accretion disk, while the influence of DM spike is typically smaller. We also illustrate that environments can leave distinctive imprints on the GW waveforms, but resolving such modifications is found to be challenging for LISA-like detectors in the near future.
\end{abstract}

\maketitle

\section{Introduction}
The first detection of gravitational waves has opened a new era of multimessenger astronomy \cite{LIGOScientific:2016aoc}. Since then, nearly one hundred gravitational-wave (GW) events have been detected by the LIGO-Virgo-KAGRA Collaboration, which are generated during the final stages of stellar-mass compact binary mergers \cite{LIGOScientific:2020iuh,LIGOScientific:2020zkf,LIGOScientific:2020aai,CanevaSantoro:2023aol}. Near future space-based GW interferometers, such as LISA and Taiji, are designed to detect GWs of lower frequencies, thus enabling the exploration of a wider range of GW events, including the inspiral and hyperbolic encounter of extreme mass-ratio binaries~\cite{Barack:2003fp,Li:2021pxf}.

Gravitational bremsstrahlung from the scattering of compact objects (COs) is a potentially important type of GW signals. Different from the gravitational waves radiated by inspiraling binaries, such bremsstrahlung radiation is burstlike, with a relatively short duration and accompanied with linear GW memory \cite{1977ApJ...216..610T,1987Natur.327..123B,PhysRevD.45.520}, which has been analyzed for binaries in vacuum using the Newtonian \cite{1977ApJ...216..610T,Capozziello:2008ra,Capozziello:2008mn, Garcia-Bellido:2017qal, Garcia-Bellido:2017knh,Caldarola:2023ipo}, post-Newtonian (PN) \cite{1992MNRAS.254..146J,  Hansen:1972jt, Majar:2010em, Berry:2012im,DeVittori:2014psa, DeVittori:2014nza, DeVittori:2014oza, Bini:2017wfr,Bini:2021jmj, Roskill:2023bmd, Cho:2018upo, Cho:2022pqy}, post-Minkowskian (PM) \cite{Jakobsen:2021smu,Riva:2022fru,DeAngelis:2023lvf} or geodesic \cite{Han:2020dql} approximation and numerical relativity~\cite{Fontbute:2024amb}. The detectability and event rate of the GW signals from compact binary hyperbolic encounters have been discussed in \cite{ Hopman:2006fc,Berry:2012en,Berry:2012im,Berry:2013poa,Capozziello:2008ra,Garcia-Bellido:2017knh,Mukherjee:2020hnm,Fan:2022wio,Wu:2023rpn,Dandapat:2023zzn,Dandapat:2024ipc}. These signals may already exist in the current observational data of ground-based GW interferometers, although searches so far have not identified them \cite{Morras:2021atg, Bini:2023gaj}. Furthermore, a large number of unresolved binary encounters can form a specific stochastic gravitational wave background~\cite{Garcia-Bellido:2021jlq, Kerachian:2023gsa}.

One particularly interesting and realistic type of scattering events is the extreme mass-ratio hyperbolic encounter (EMRHE) between stellar-mass COs and the supermassive black holes (SMBHs) at the galaxy centers \cite{Berry:2012im}. In the case of Sgr A*, the estimated detectable event rate of EMRHEs is about $0.1\textendash 1\,\text{yr}^{-1}$ \cite{Hopman:2006fc,Berry:2013ara}, which is much higher than the estimation $10^{-6}\textendash 10^{-5}\,\text{yr}^{-1}$ \cite{Amaro-Seoane:2010dzj,Amaro-Seoane:2012lgq,Babak:2017tow,2022hgwa.bookE..17A} for the capture event rate of extreme mass-ratio inspirals (EMRIs). The vicinity of a SMBH is likely to host rich and important physical processes that are beyond the reach of direct observations. An indirect messenger is provided by the gravitational waves emitted by small COs orbiting the central BH, whose motion can be affected by the black hole environments gravitationally, thus leaving imprints on their GW waveforms. This possibility has been extensively studied for the inspiraling extreme or intermediate mass-ratio binaries~\cite{Kocsis:2011dr,PhysRevLett.107.171103, Bertone:2019irm, PhysRevD.103.023015, Becker:2021ivq, Cardoso:2021wlq,Cardoso:2022whc,Becker:2022wlo, Yue:2017iwc, Yue:2018vtk, Barausse:2014tra, Cole:2022yzw,Rahman:2023sof, Bertone:2024rxe, Yue:2019ozq, Yue:2024xhf, Zhang:2024hrq, Montalvo:2024iwq, Li:2021pxf, Zhang:2024csc, Zhao:2024bpp,Speeney:2024mas,Duque:2023seg,Kadota:2023wlm,kim2024gravitationalwaveduetresonating}. A similar situation could occur in the case of unbound orbits. But so far only the impact of DM spike has been discussed in the context of intermediate mass-ratio hyperbolic encounters~\cite{AbhishekChowdhuri:2023rfv}.

In this paper, we study the gravitational effects of two types of environments - the accretion disk and the DM spike - on the orbital evolution and GW waveforms of EMRHEs around SMBHs. EMRHEs are affected by the BH environment mainly through its gravitational potential, whose effect in some cases turn out to be more significant than the relativistic corrections. The influence of DM spike is found to be typically weak, while the presence of accretion disk can have potentially large impacts. We also examine the detectability of gravitational wave signals from EMRBHE events in the centers of Milky Way and the nearby galaxy M31, as well as the distinguishability of environmental imprints in the waveforms. We find that events showing distinctive waveform modifications due to the environmental influences are generally beyond the sensitivity range of upcoming LISA-like detectors, while the environmental signatures can be challenging to resolve in the waveforms of more detectable events.

The structure of our paper is as follows. In Sec.~\ref{sec2}, we present a general framework for the computation of orbital evolution during the hyperbolic encounter. The physical models of DM spike and accretion disk are reviewed in Sec.~\ref{sec3}, where we also derive the corresponding environmental forces. The orbital evolution in the presence of environments are investigated concretely in Sec.~\ref{sec4} and in Sec.~\ref{sec5}, we compute and discuss the associated GW waveforms. We give our conclusions in Sec.~\ref{sec6}. Throughout this paper, we use the natural units $G = c = \hbar = 1$.

\section{Orbital Parametrization and Dynamics}\label{sec2}
\subsection{Initial condition}

\begin{figure}[t]
    \centering
    \includegraphics[width=0.45\textwidth]{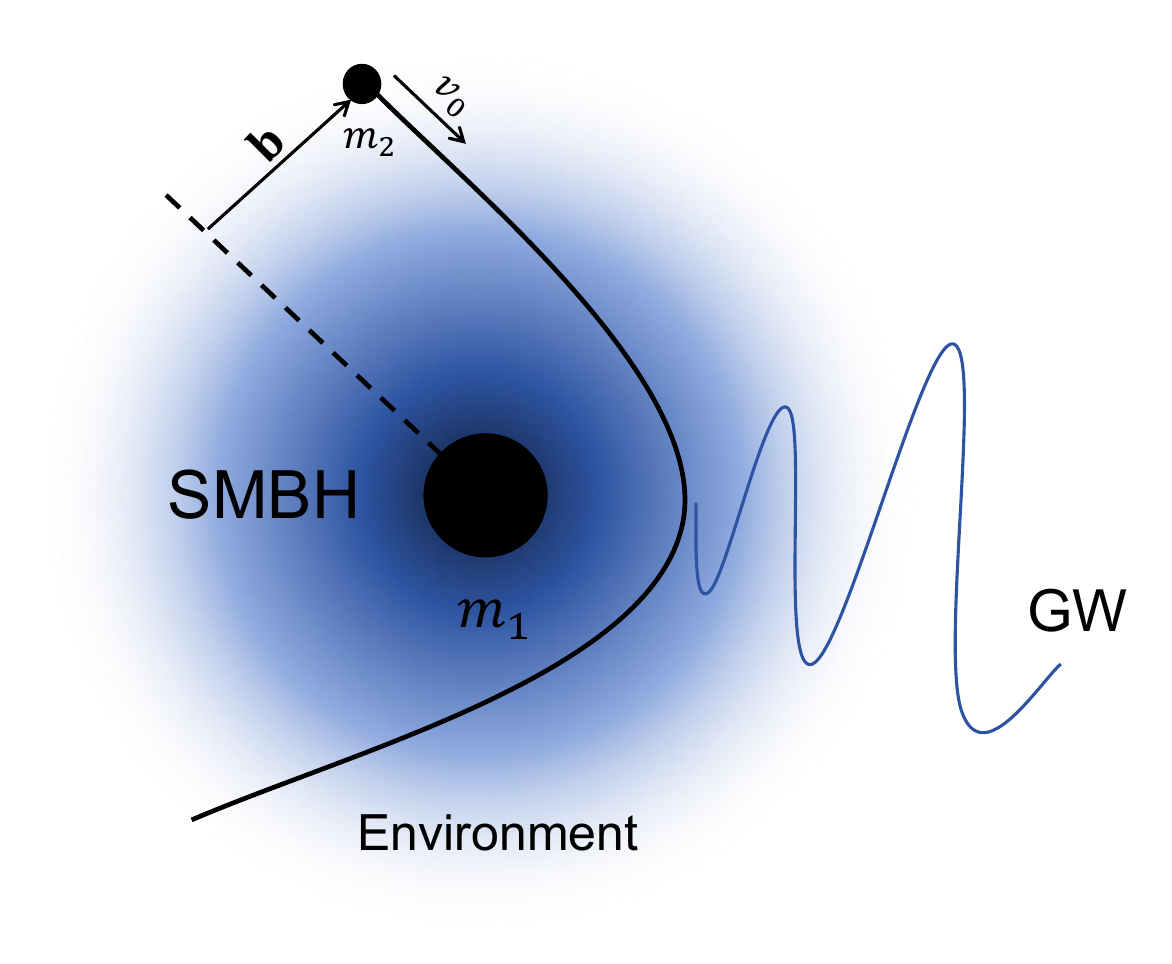}
    \caption{Sketch of an EMRHE in the presence of BH environment. A burstlike GW is emitted during this process.}
    \label{fig:diagram}
\end{figure}

\begin{figure*}[htbp]
	\centering
	\includegraphics[width=0.9\textwidth]{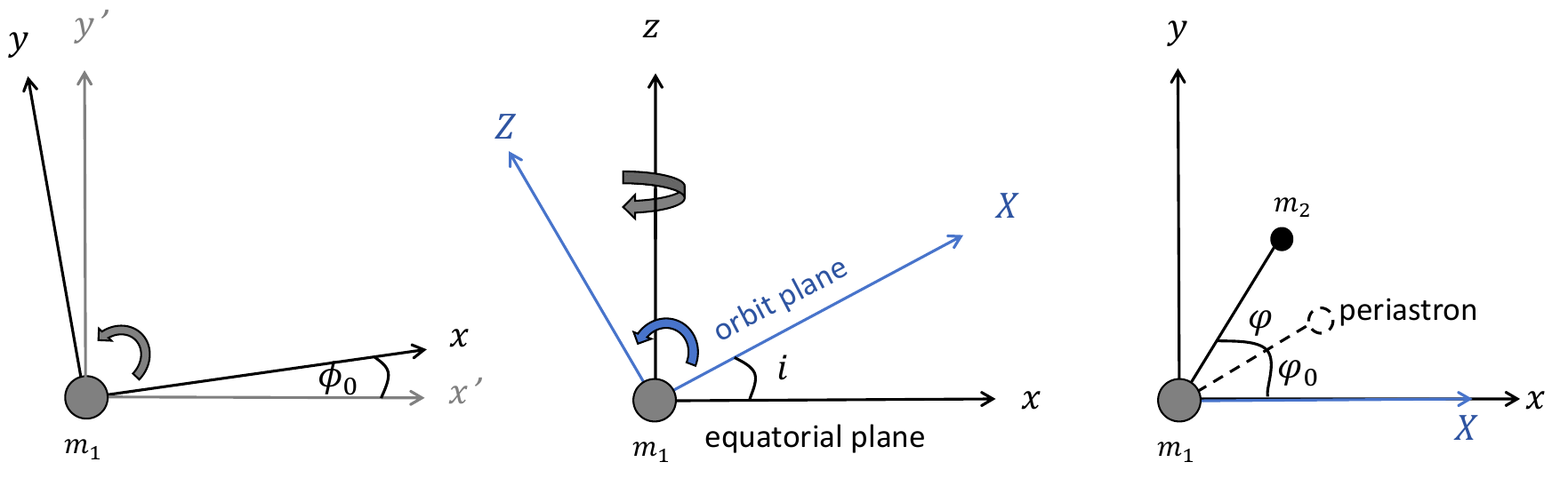}
    \caption{The definitions of three Cartesian coordinate systems ($x'y'z'$, $xyz$, $XYZ$) and the angular osculating orbital elements $\{\phi_0,i,\varphi_0\}$. The central BH lies in the coordinate origin with its spin parallel to the $z'=z$ axis. The osculating orbit lies on the $XY$ plane, with $Y=y$. The $x'y'$ plane is the reference plane.}
    \label{fig:coor}
\end{figure*}

We consider a small compact object with mass $m_2$ scattered by a massive central BH with mass $m_1$, as illustrated in Fig.~\ref{fig:diagram}. The position vectors of the two bodies are denoted by $\mathbf{r}_{1,2}$ in the center-of-mass (COM) frame, with the relative motion given by $\mathbf{r} = \mathbf{r}_2 - \mathbf{r}_1$ and $\mathbf{v} =\dot{\mathbf{r}}$. At the Newtonian order, this two-body scattering corresponds to a hyperbolic orbit which can be parametrized by six orbital elements $\{a,e,\varphi_0,\phi_0,i,t_0\}$ with respect to a fixed reference plane (see Fig.~\ref{fig:coor} and also Appendix~\ref{appendix_analytical}), here $a$ is the semimajor axis, $e$ the eccentricity, $\varphi_0$ the argument of periastron on the orbital plane, $\phi_0-\pi/2$ the longitude of the ascending node and $i$ the inclination angle of the orbital plane relative to the reference plane.

Under perturbations, the true orbit will deviate from the Keplerian one\footnote{If the cumulative dissipation is sufficiently strong, the binary orbit can become bound, resulting in a so called dynamical capture \cite{Albanesi:2024xus}. The dynamical captures due to GW damping alone would happen for sufficiently small impact parameter or incident velocity \cite{Mouri:2002mc}. We focus instead on the regime of fast hyperbolic encounter (with large enough $v_0$, or equivalently, large enough $e_0$ for given $b$), during which such capture would not happen, and the CO does not plunge into the central BH.}, but it can always be described with an evolving osculating Keplerian orbit, with the position and velocity matched with the true orbit at any given moment. We shall assume that the initial velocity $\mathbf{v}_0$ is small enough and the environment has a spatially compact support, then at sufficiently large distance from the central BH, the motion should be well approximated by a vacuum Keplerian orbit, and the initial state of this scattering process can be specified by the impact parameter vector $\mathbf{b}=b\,(\mathbf{e}_r\times\mathbf{e}_Z)$ and the relative velocity $\mathbf{v}_0$ in the asymptotic past, corresponding to an initial osculating orbit with \cite{bate1971fundamentals} $i_0\equiv i(t=0)$,
\begin{align}
e_0 & \equiv e(t=0)=\sqrt{1+\frac{b^2 v_{0}^{4}}{M^2}},
\\
a_0 & \equiv a(t=0)=\frac{b}{\sqrt{e_0^2-1}},
\end{align}
where $M=m_1+m_2$ is the total mass.

It is convenient to perform the numerical computation in spherical coordinates $(r,\theta,\phi)$, with
\begin{align}
\mathbf{r}&=r\,\mathbf{e}_r,
\\
\mathbf{v}&=\dot{r}\,\mathbf{e}_r+r\,\dot{\theta}\,\mathbf{e}_\theta+r\,\sin{\theta}\,\dot{\phi}\,\mathbf{e}_\phi.
\end{align}
The Cartesian components of the initial position $\mathbf{r}(t=0)$ and velocity $\mathbf{v}(t=0)$ in the fixed $x'y'z'$ coordinate system (defined in Fig.~\ref{fig:coor}) are fixed by the initial osculating elements (see Appendix~\ref{elements} for details), whose spherical components are
\begin{align}
    r&=\frac{a_0(e_0^2-1)}{1+e_0\cos{\varphi(t=0)}},\\
    \theta&=\arccos{(z'/r)},\\
    \phi&=\arctan{(y'/x')},
\end{align}
and
\begin{align}
v_r&=v_{x'}\sin{\theta}\cos{\phi}+v_{y'}\sin{\theta}\sin{\phi}+v_{z'}\cos{\theta},
\\
v_\theta&=v_{x'}\cos{\theta}\cos{\phi}+v_{y'}\cos{\theta}\sin{\phi}-v_{z'}\sin{\theta},
\\
v_\phi&=-v_{x'}\sin{\phi}+v_{y'}\cos{\phi},
\end{align}
with $\dot r = v_{r}$, $\dot\theta = v_\theta/{r}$ and $\dot\phi = v_{\phi}/(r \sin{\theta})$. We set $\varphi_0(t=0)=\phi_0(t=0)=0$ and $\varphi(t=0)=-0.99(\pi-\arctan \sqrt{e_0^2-1})$. In all cases we considered, this generates a nearly Keplerian initial orbit.

\subsection{Equation of motion}
To account for the PN corrections and environmental effects, we incorporate these as an additional acceleration term $\mathbf{F}$ in the Newtonian equation of the binary system: $\ddot{\mathbf{r}}=-(M/r^2)\,\mathbf{e}_r+\mathbf{F}$. The explicit, componentwise form is given by
\begin{align}
    \ddot r&=-\frac{M}{r^2}+r(\dot\theta)^2+r\sin^2{\theta}\,(\dot\phi)^2+ F_r, \label{EOM_1}
\\
    \ddot\phi&=-\frac{2}{r}\dot\phi\,\dot{r}-2\,\dot\phi\,\dot\theta\,\cot{\theta}+\frac{F_\phi}{r\sin{\theta}}, \label{EOM_2}
\\
    \ddot\theta&=-\frac{2\,\dot r\, \dot\theta}{r}+(\dot\phi)^2\cos{\theta}\,\sin{\theta}+\frac{F_\theta}{r}, \label{EOM_3}
\end{align}
with $F_{r,\phi,\theta}=\mathbf{F}\cdot\mathbf{e}_{r,\phi,\theta}$. The relative acceleration due to 1 PN and 2.5 PN corrections (being the leading-order terms in the conservative and dissipative sectors, respectively) in the harmonic gauge are as follows \cite{Blanchet:2013haa}
\begin{align}
    \frac{\mathbf{F}^{(\mathrm{1PN})}}{M/r^2}=& \left[(4+2\nu)\,\frac{M}{r}-(1+3\nu)\,v^{2}+\frac{3}{2}\nu\,(\dot{r})^{2}\right]\mathbf{e}_r \nonumber
	\\
	&+(4-2\nu)\,\dot{r}\,\mathbf{v}, \label{F_1PN}
 \\
    \frac{\mathbf{F}^{(2.5\mathrm{PN})}}{M^2/r^3}=& \frac{24\nu}{5}\left[\left(v^{2}+\frac{17M}{9r}\right)\dot{r}\,\mathbf{e}_r  
	-\left(\frac{v^{2}}{3}+\frac{M}{r}\right)\mathbf{v}\right],\label{F_2.5PN}
\end{align}
where $v=|\mathbf{v}|$, $\nu=\mu/M$ is the symmetric mass ratio and $\mu = m_1 m_2/M$ is the reduced mass. In the nonvacuum scenario, the radiation damping given by Eq.~\eqref{F_2.5PN} can also be approximately applied if the perturbation to the Keplerian motion is small. We do not consider higher-order PN corrections, because our focus is on the moderately relativistic regime, which should also be more relevant for the astrophysical scenarios. Once the acceleration $\mathbf{F}$ is specified, we numerically solve Eqs.~\eqref{EOM_1}, \eqref{EOM_2}, \eqref{EOM_3} to determine the orbit $\mathbf{r}(t)$.

\section{Black Hole Environments}\label{sec3}
From its gravitational interaction with the ambient matter, SMBH in the galaxy center is possibly surrounded by a dense environment, such as the baryonic accretion disk and density spike of particle DM. In this section, we review the modeling of these two types of environments and the resultant gravitational effects on the motion of CO, including the dissipative dynamical friction and the environmental gravitational potential. In the regime of fast hyperbolic encounter, the latter effect will be more important, since the dissipative force is typically much weaker than the conservative one.

\begin{figure*}[ht]
	\centering
	\includegraphics[width=0.48\textwidth]{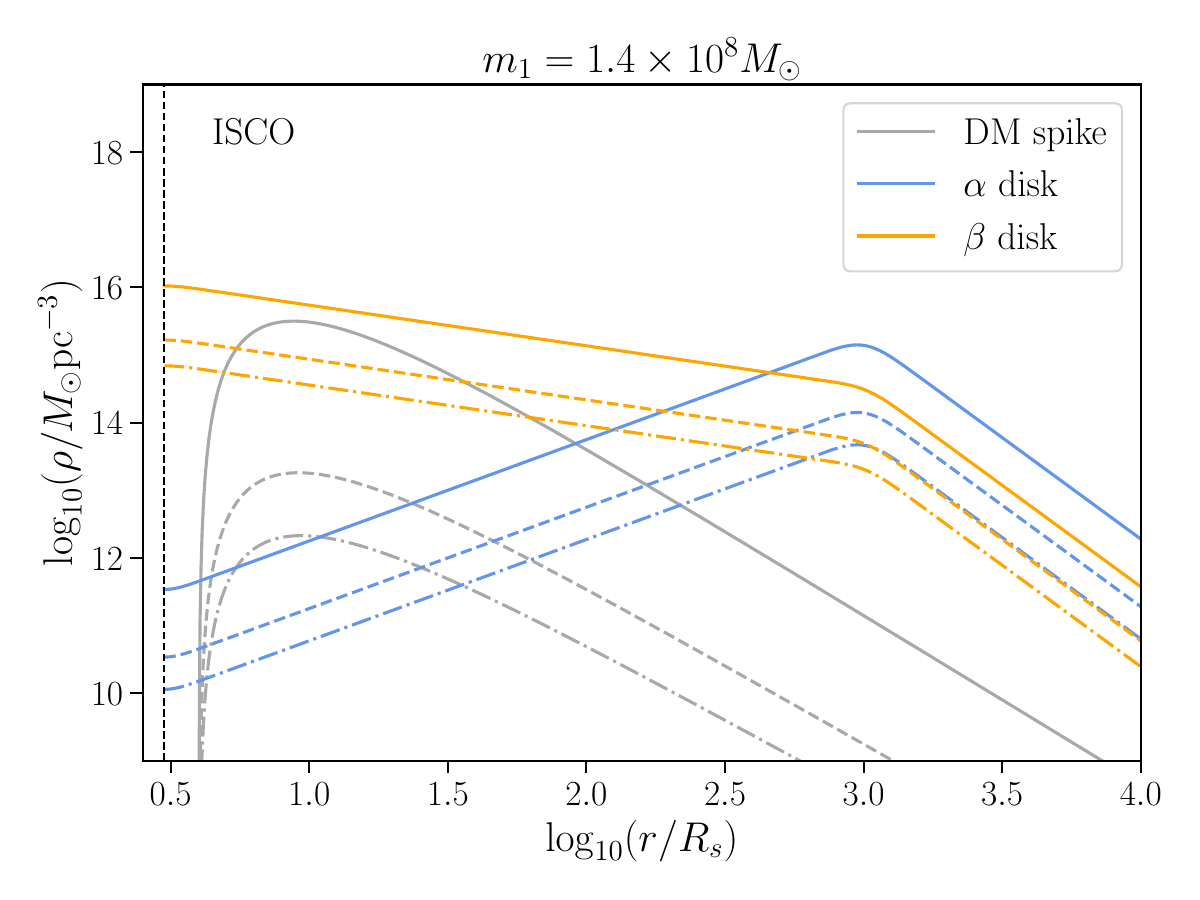}
        \;\;\includegraphics[width=0.48\textwidth]{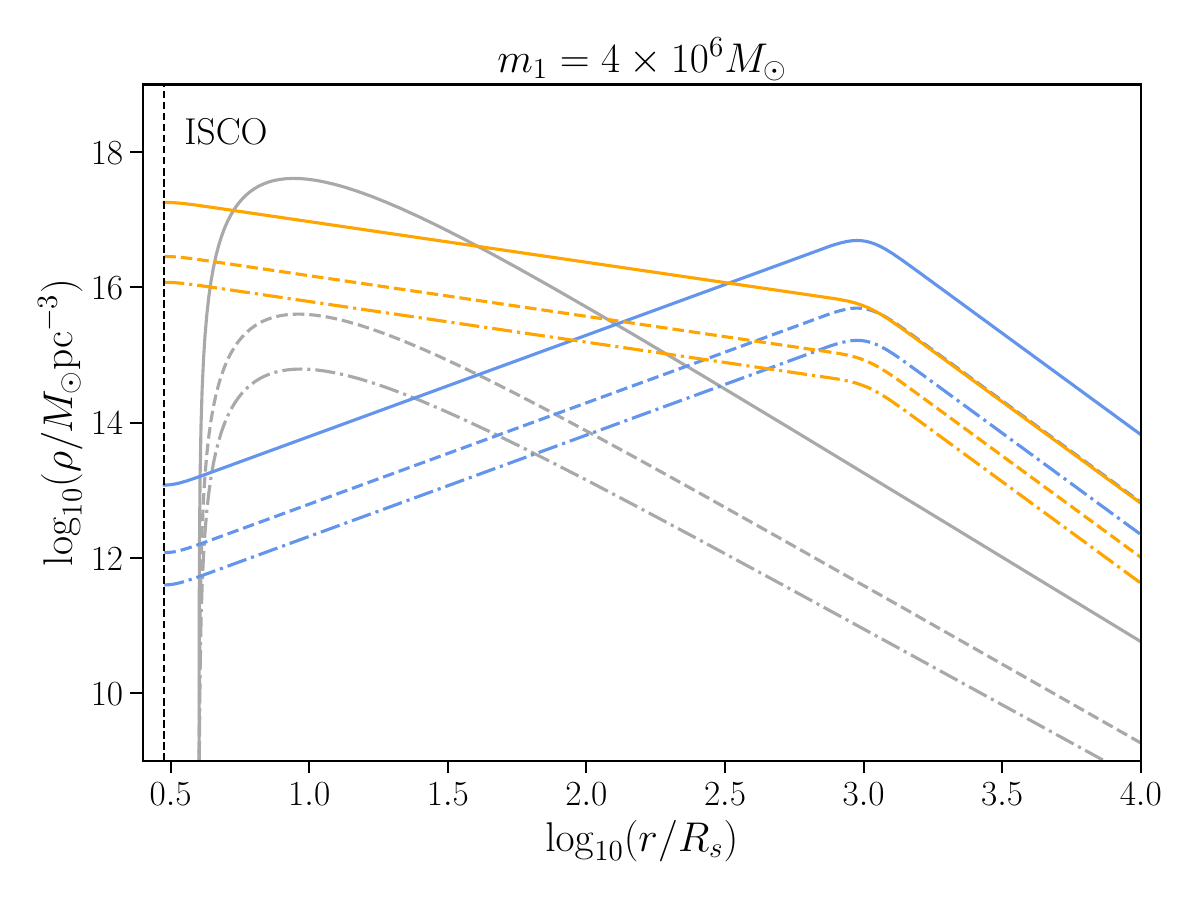}
	\caption{The mass density distribution of accretion disk and DM spike around a SMBH with mass $m_1=1.4\times 10^8 \,M_\odot$ (left panel) or $4\times10^6 \,M_\odot$ (right panel). The solid, dashed and dot-dashed curves correspond to the viscosity coefficient $\alpha=0.01,0.1,0.3$ for the disk, or the power-law index $\gamma=2,1,0$ for the DM spike, respectively.}
\label{fig:density}
\end{figure*}

\subsection{DM spike}
If the dark matter is composed by massive particles such as weakly-interacting massive particles (WIMPs), the past growth of the central black hole might have enhanced the density of these DM particles at the galaxy center, which can be estimated as follows. We assume the initial density profile of the DM halo is
\begin{equation}
\rho(r)\approx\rho_0{(r/r_0)}^{-\gamma}.\label{initial_spike}
\end{equation}
If the central BH forms and grows adiabatically through accretion of the ambient DM particles, a density spike would eventually form. The density distribution of this DM spike is given by~\cite{Gondolo:1999ef, Sadeghian:2013laa, Hannuksela:2019vip, Kavanagh:2020cfn}
\begin{equation}\label{spike_density}
    {\rho}(r)={\rho}_\text{sp}(1-4R_\text{s}/r)^{3}(r/r_\text{sp})^{-\xi},
\end{equation}
and $\rho(r<4R_\text{s})=0$, where $R_\text{s}=2 m_1$ is the Schwarzschild radius of the central BH, the parameter $\xi=(9-2\gamma)/(4-\gamma)>2$ ($\xi$ increases with $\gamma$, e.g., for $0 < \gamma < 2$, $\xi$ falls between 2.25 and 2.5). Here we neglect the possible modifications to this profile due to DM annihilation \cite{Gondolo:1999ef}, interaction with the surrounding stellar cluster \cite{PhysRevLett.93.061302} as well as alternative formation scenarios \cite{Ullio:2001fb}. The parameters $\rho_\text{sp}$ and $r_\text{sp}$ depend on the SMBH mass $m_1$ and the initial density distribution \eqref{initial_spike}, the concrete estimation is detailed in Appendix \ref{app_dm}. In Fig.~\ref{fig:density} we show in gray curve the density distribution of DM spike for different values of $\gamma$.

\subsubsection{Dynamical friction}
Due to gravitational scattering with DM particles, the CO experiences a dynamical friction (DF) as it travels in the spike. This DF can be estimated by the classical formula of Chandrasekhar at the Newtonian order\footnote{We neglect the relativistic correction to the momentum transfer \cite{1969ApJ...155..687L,Chiari:2022kas} as well as the contribution from DM particles moving faster than the CO \cite{Dosopoulou:2023umg}.} \cite{Chandrasekhar:1943ys}
\begin{equation}
    \mathbf{F}^\text{(DF)}=-\frac{4\pi{m_2}\,\rho_\text{DM}\,\epsilon(v)\,\ln\Lambda}{v^3}\mathbf{v},
\end{equation}
where $\epsilon(v)$ is the fraction of DM particles with velocity smaller than $v$, $\ln \Lambda$ is the Coulomb logarithm given by \cite{Kavanagh:2020cfn} $\ln\Lambda=\ln\left({b_{\text{max}}}/{b_{\text{min}}}\right)\approx -\frac{1}{2}\ln\left({m_1}/{m_2}\right)$, with $b_{\text{max (min)}}$ being the maximum (minimum) impact parameter of DM particles relative to the CO. E.g., for $m_2=10\,M_\odot$ and $m_1=10^8 \,M_\odot$ ($10^6 M_\odot$), this gives $\ln \Lambda \approx 8$ (6). For simplicity, we shall use the approximation $\epsilon(v) \approx 1$.

\subsubsection{Gravitational potential}
\begin{figure}[t]
    \centering
    \includegraphics[width=0.48\textwidth]{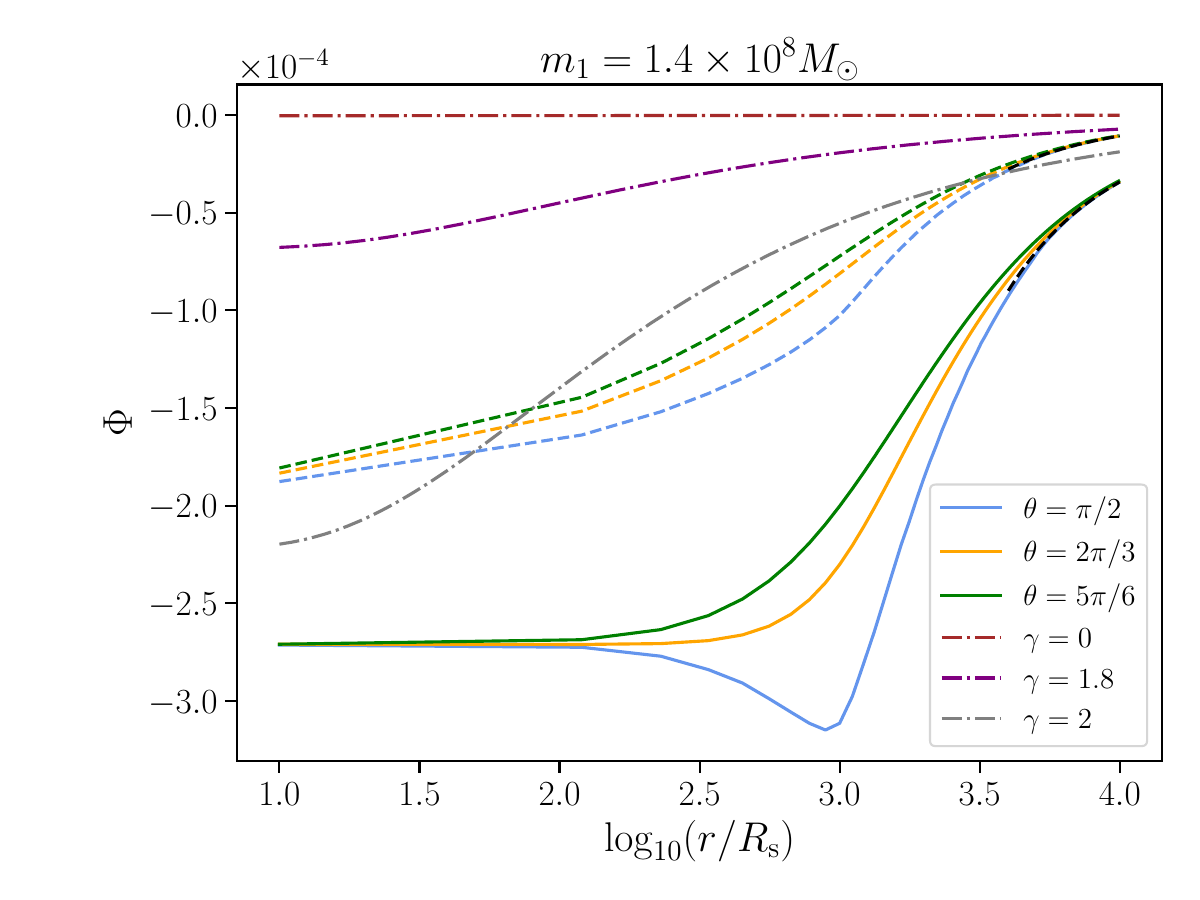}
    \caption{Environmental gravitational potential of a SMBH with mass $m_1=1.4\times 10^8\,M_\odot$. The solid, dashed and dot-dashed curves correspond to the cases of $\alpha$ disk, $\beta$ disk and DM spike, respectively. The black dashed curve shows the monopole + quadrupole approximation \eqref{quadrupole}. For the accretion disk, we set $\alpha=0.02$ and $l_\text{E}=0.1$.}
    \label{fig:phi}
\end{figure}

The density distribution $\rho(\mathbf{r})=\rho(r)$ of the considered DM spike is spherically symmetrical, correspondingly the solution of Poisson equation $\nabla^2\Phi=4\pi \rho$ :
\begin{equation}
\begin{aligned}
\Phi(r)=&-4\pi \left[\frac{1}{r}\int_0^r\rho_(r')\,{r'}^2dr'+\int_r^\infty\rho(r')\,r'dr'\right]\\
=&\,4\pi \rho_{\text{sp}} \left(\frac{r_{\text{sp}}}{r}\right)^{\xi} \bigg\{
\frac{r^2}{\xi - 3}
- \frac{r^2 }{\xi - 2}
+ \frac{192 m_1^2}{(\xi - 1)\, \xi}\\
&- \frac{512 m_1^3 }{r\, \xi}\bigg[
    \frac{1}{\xi + 1}
    + \frac{6\left({r}/{8m_1}\right)^{\xi}}{(\xi-3)(\xi-2)(\xi-1)}\bigg]\\
&- \frac{24 m_1 r}{(\xi-2)(\xi-1)}\bigg\},   
\end{aligned}
\end{equation}
and the gravitational acceleration is
\begin{equation}
    \mathbf{F}^\text{(GP)}=-\nabla \Phi=-(\partial_r \Phi)\, \mathbf{e}_r=F_{r}^\text{(GP)}\,\mathbf{e}_r,
\end{equation}
where
\begin{equation}
\begin{aligned}
    F_{r}^\text{(GP)}=&-4\pi\frac{\rho_\text{sp}}{r^2}\left({\frac{r_\text{sp}}{r}}\right)^{\xi} \\
    &\times
    \left\{\frac{r^3}{3-\xi}+\frac{24 m_1 r^2}{\xi-2}-\frac{192 {m^2_1}r}{\xi-1}+\frac{512 m_1^3}{\xi} + \right. \\
    &\left. 
    \frac{3072 {m^3_1}}{\xi}\left[(\xi+1)+\frac{ r^{\xi}(8 m_1)^{-\xi}}{(\xi-3)(\xi-2)(\xi-1)}\right]\right\}.
\end{aligned}
\end{equation}
The gravitational potential of DM spike is shown in Fig.~\ref{fig:phi} as dot-dashed curve.
\subsection{Accretion disk}
We consider a stationary accretion disk that is geometrically thin, optically thick and radiatively efficient, with opacity primarily determined by the electron scattering, known as the standard thin disk model \cite{1973IAUS...55..155S,frank2002accretion,2020ApJ...897..142S,Liu:2024ekz}. The disk temperature is sufficiently high for complete ionization, and the gravitational potential is dominated by the central black hole.

In order to study generic orbits with finite inclination to the disk plane, a three-dimensional model for the gravitational potential of the disk is needed. To construct a reasonable and complete density profile, we divide the disk into the inner ($r<R_\text{Q}=10^3R_\text{s}$) and outer ($R_\text{Q}<r<10 R_\text{Q}$) regions. We use the standard  ($\alpha$ and $\beta$) disk model \cite{2004ApJ...608..108G} with a constant scale height to describe the inner region, and the self-gravitating disk model \cite{Goodman:2003sf} to describe the outer region  (see Appendix~\ref{disk} for details). The resultant density profiles for the $\alpha$ and $\beta$ disks are
\begin{equation}
\begin{aligned}
    \rho_{\beta}&= 2.52\times10^{-7}\,\left(\frac{\alpha}{0.3}\right)^{-4/5}\left(\frac{m_1}{10^8M_\odot}\right)^{-4/5}\\
    &
    \times
    \begin{cases}
        \exp{\left(-\frac{Z^2}{2{H^2_{\text{in}}}}\right)}\left(\frac{R}{R_\text{Q}}\right)^{-3/5}\text{g}\,\text{cm}^{-3}, & R\leq R_\text{Q}\\
        \exp{\left(-\frac{Z^2}{2{H^2_{\text{out}}}}\right)}\left(\frac{R}{R_\text{Q}}\right)^{-3}\text{g}\,\text{cm}^{-3}, &R> R_\text{Q}
    \end{cases}
\\
    \rho_{\alpha}&= 5.51\times10^{-8}\,\left(\frac{\alpha}{0.3}\right)^{-1}\left(\frac{m_1}{10^8M_\odot}\right)^{-1}\\
    &
    \times
    \begin{cases}
        \exp{\left(-\frac{Z^2}{2{H^2_{\text{in}}}}\right)}\left(\frac{R}{R_\text{Q}}\right)^{3/2}\text{g}\,\text{cm}^{-3},
        &R\leq R_\text{Q}\\
        \exp{\left(-\frac{Z^2}{2{H^2_{\text{out}}}}\right)}\left(\frac{R}{R_\text{Q}}\right)^{-3}\,\text{g}\,\text{cm}^{-3},
        &R> R_\text{Q}
    \end{cases}
\end{aligned}
\end{equation}
where $\alpha$ is the viscosity coefficient, $H_{\text{in(out)}}$ the scale height of the inner (outer) region, and $(R,Z)$ the cylindrical coordinates. In Fig.~\ref{fig:density} we show the density distribution of accretion disk for different values of $\alpha$.

\subsubsection{Dynamical friction}
The CO could experience several types of dissipative forces in the accretion disk, such as DF, migration, accretion, azimuthal and radial winds \cite{Kocsis:2011dr,2024arXiv241103436D}. For simplicity, here we consider only the effect of DF. In the supersonic regime, the DF in a gaseous medium can be described by the Ostriker model \cite{1999ApJ...513..252O}:
\begin{equation}
    \mathbf{F}^{(\text{DF})}=-\frac{4\pi m_2\rho}{{v_\text{rel}}^3}\ln\left(\frac{r_\text{max}}{r_\text{min}}\right)\,\textbf{v}_\text{rel},
\end{equation}
where $\textbf{v}_\text{rel}$ is the velocity of CO relative to the gas, and $r_\text{max}$ and $r_\text{min}$ correspond respectively to the effective linear sizes of the surrounding medium and the perturbing object. In all cases considered below, the Mach number $v_\text{rel}/c_\text{s} > 10$ and the supersonic condition is fully satisfied. We choose \cite{Vicente:2019ilr} $r_\text{max}=H$ and $r_\text{min}=R_\text{HL}$, here $R_\text{HL}=2\,m_2/v_\text{rel}^2$ is the Hoyle-Lyttleton radius \cite{2025arXiv250208288A} below which the gas is gravitationally bound to the CO and thus does not contribute to the momentum transfer.

\begin{figure*}[htbp]
\centering
	\includegraphics[width=1\textwidth]{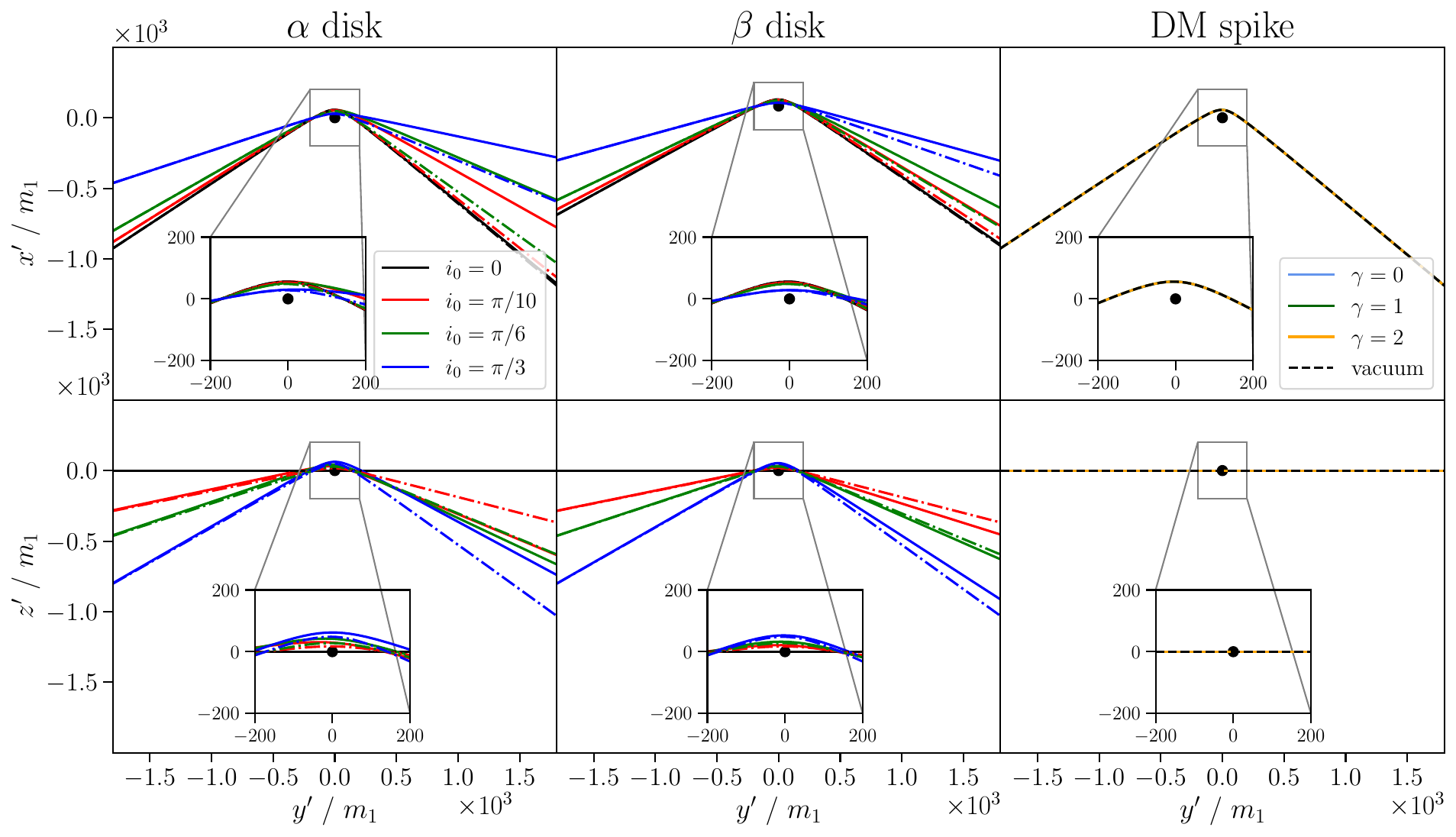}
        \includegraphics[width=1\textwidth]{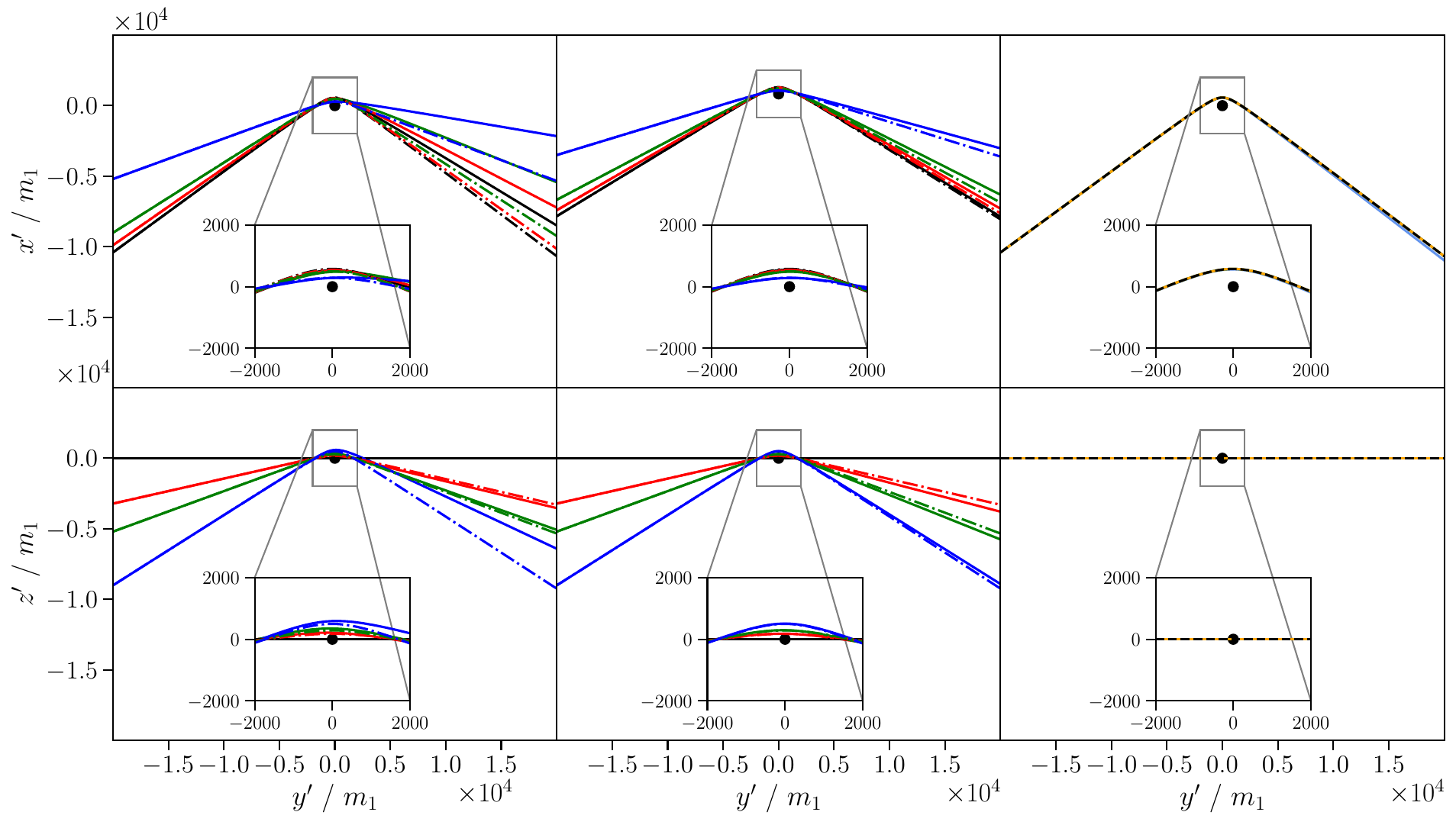}
	\caption{The projected trajectories of hyperbolic encounters on the $x'y'$ and $y'z'$ planes, with the impact parameter $b=50\,R_\text{s}$ (upper panel) or $500\,R_\text{s}$ (lower panel), $e_0 = 2$, $m_1 = 1.4\times 10^8 \, M_\odot$, $m_2 = 10 \, M_\odot$, $\varphi_0(t=0)=\phi_0(t=0)=0$ and different initial orbital inclinations. The viscosity coefficient of the accretion disk is set to be $\alpha = 0.02$. The solid (dot-dashed) curves correspond to the results in the presence (absence) of accretion disks.}
\label{fig:orbit_incli}
\end{figure*}
\subsubsection{Gravitational potential}
The density distribution $\rho(\mathbf{r})=\rho(R,Z)$ of the accretion disk is cylindrically symmetrical, correspondingly the solution to Poisson equation $\nabla^2\Phi=4\pi \rho$ is \cite{binney2008galactic}
\begin{equation}
\begin{aligned}
    \Phi(R,Z)&=\int_{-\infty}^{\infty}dZ'\int_{R_{\text{min}}}^{R_{\text{max}}}dR'\,\frac{-4 R'\rho(R',Z')}{\sqrt{(R+R')^2+(Z-Z')^2}}
    \\
    &\qquad \times K\left(\frac{4\,RR'}{(R+R')^2+(Z-Z')^2}\right).
\end{aligned}
\end{equation}
The spherical components of the resultant gravitational acceleration are given by
\begin{widetext}
\begin{align}
    F_{r}^{(\text{GP})}(r,\theta)&=\int_{-\infty}^{\infty}dZ'\int_{R_\text{min}}^{R_\text{max}}dR'\,
    \frac{2R'\rho(R',Z')}{r\sqrt{Y_{+}}}\left[\frac{(R')^2-r^2+(Z')^2}{Y_{-}}E(
    \zeta
    )-K(\zeta)\right],
    \\[1em]
    F_{\theta}^{(\text{GP})}(r,\theta)&=\int_{-\infty}^{\infty}dZ'\int_{R_\text{min}}^{R_\text{max}}dR'\,\frac{2R'\rho(R',Z')\cot{\theta} }{r^2\sqrt{Y_{+}}}\left\{ \frac{\cot{\theta}\,[(R')^2+r^2+(Z')^2]-2 Z'r\cos{\theta}}{Y_{-}}E(\zeta)
    -K(\zeta)\right\},
\end{align}
\end{widetext}
with $Y_{\pm}=(R')^2+r^2+(Z')^2\pm2 R'r \sin{\theta}-2 Z' r\cos{\theta}$ and $\zeta={4 R' r \sin{\theta}}/[{(R'+r \sin{\theta})^2+(Z'-r \cos{\theta})^2}]$, here $K$ ($E$) is the elliptic integral of the first (second) kind. We choose $R_{\text{min}}$ to be the ISCO radius and $R_{\text{max}}$ to be the outer radius of the disk. The gravitational potential of $\alpha\,(\beta)$ disk is shown in Fig.~\ref{fig:phi} as solid (dashed) curve.

\section{Environmental Effects on the Orbital Evolution}\label{sec4}

\begin{figure*}[htbp]
	\centering
	\includegraphics[width=0.9\textwidth]{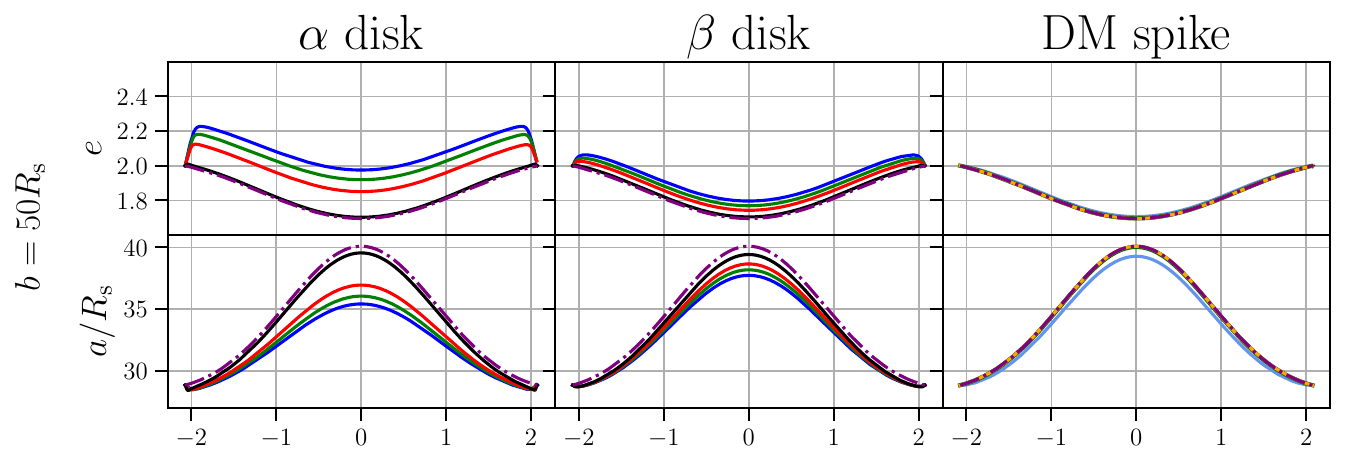}
\qquad
    \includegraphics[width=0.9\textwidth]{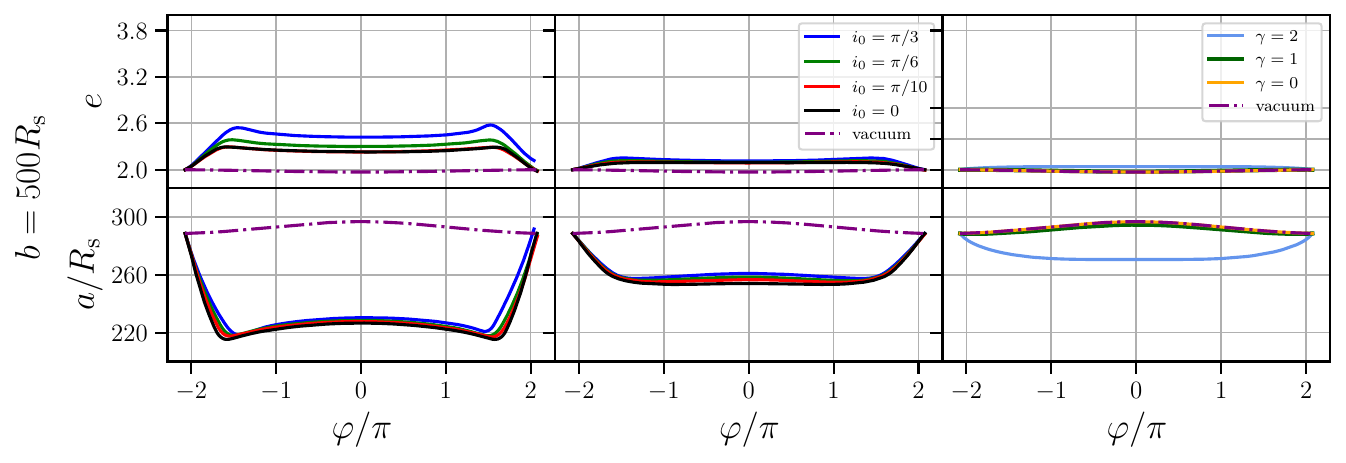}
	\caption{Evolution of eccentricity $e$ and semimajor axis $a$ during the hyperbolic encounter, with $m_1=1.4\times10^8 \,M_{\odot}$, $m_2=10 \,M_{\odot}$, $e_0=2$ and the impact parameter $b=50\,R_\text{s}$ (upper panel) or $500\,R_\text{s}$ (lower panel). The viscosity coefficient of the accretion disk is set to be $\alpha = 0.02$.}
\label{fig:orbit_para1}
\end{figure*}

\begin{figure}[hbt!]
    \centering
    \includegraphics[width=0.4\textwidth]{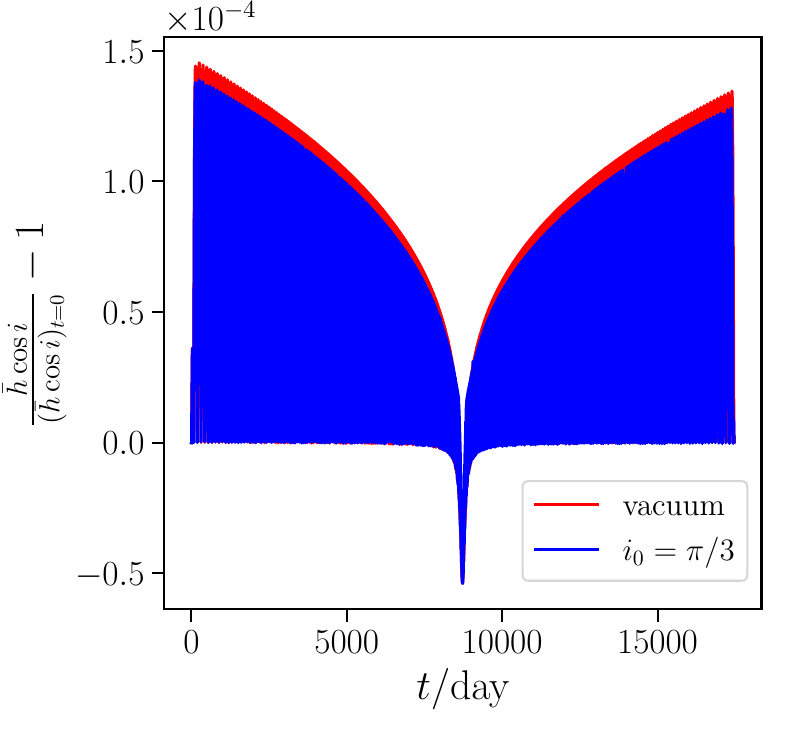}
    \caption{Evolution of the projected orbital angular momentum $\bar{h}\cos i$ during the hyperbolic encounter in the presence of $\alpha$-disk, with $m_1=1.4\times10^8 \,M_{\odot}$, $m_2=10 \,M_{\odot}$, $e_0=2$ and $b=500\,R_\text{s}$.}
    \label{fig:hz}
\end{figure}

\begin{figure*}[htbp]
	\centering
	\includegraphics[width=0.9\textwidth]{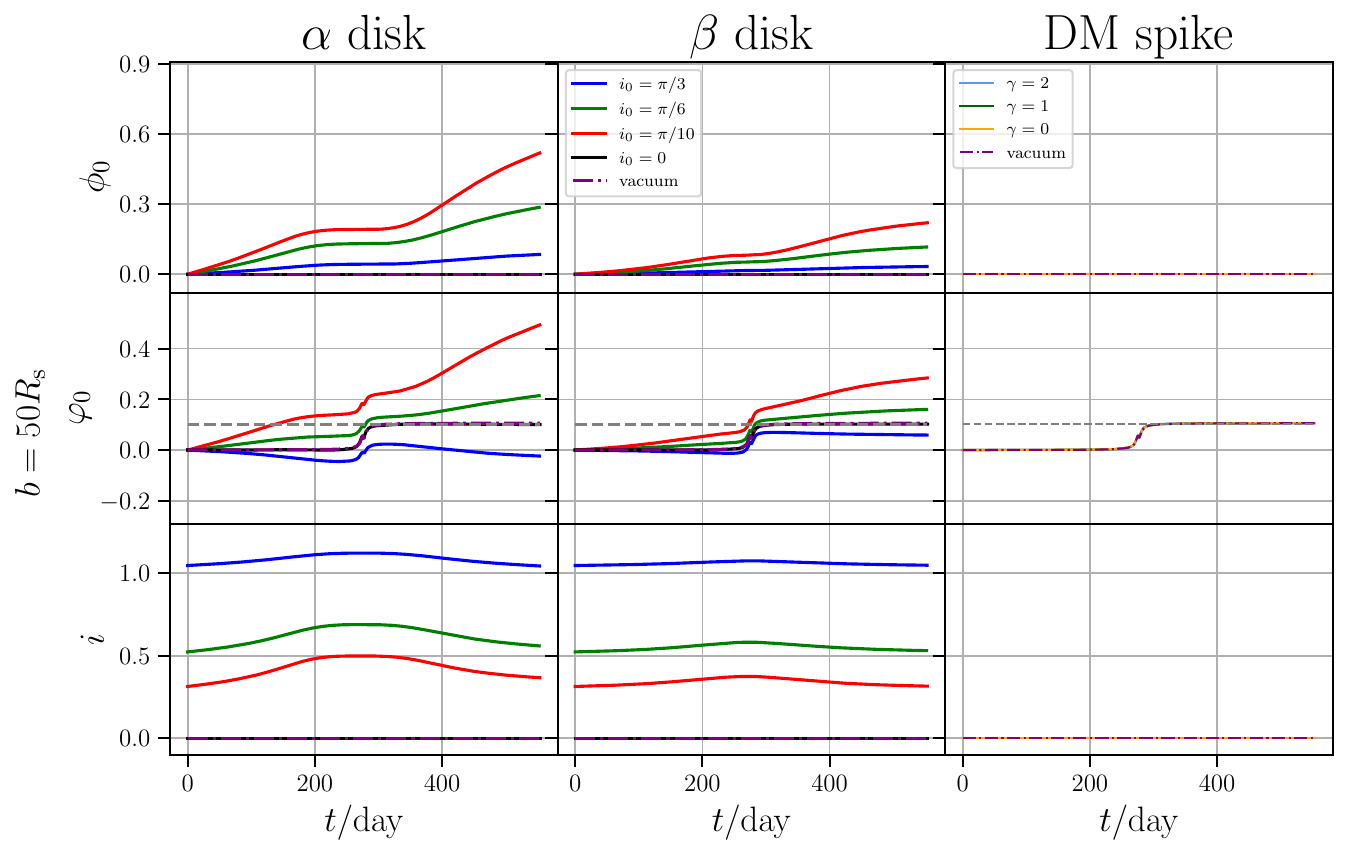}
\qquad
    \includegraphics[width=0.9\textwidth]{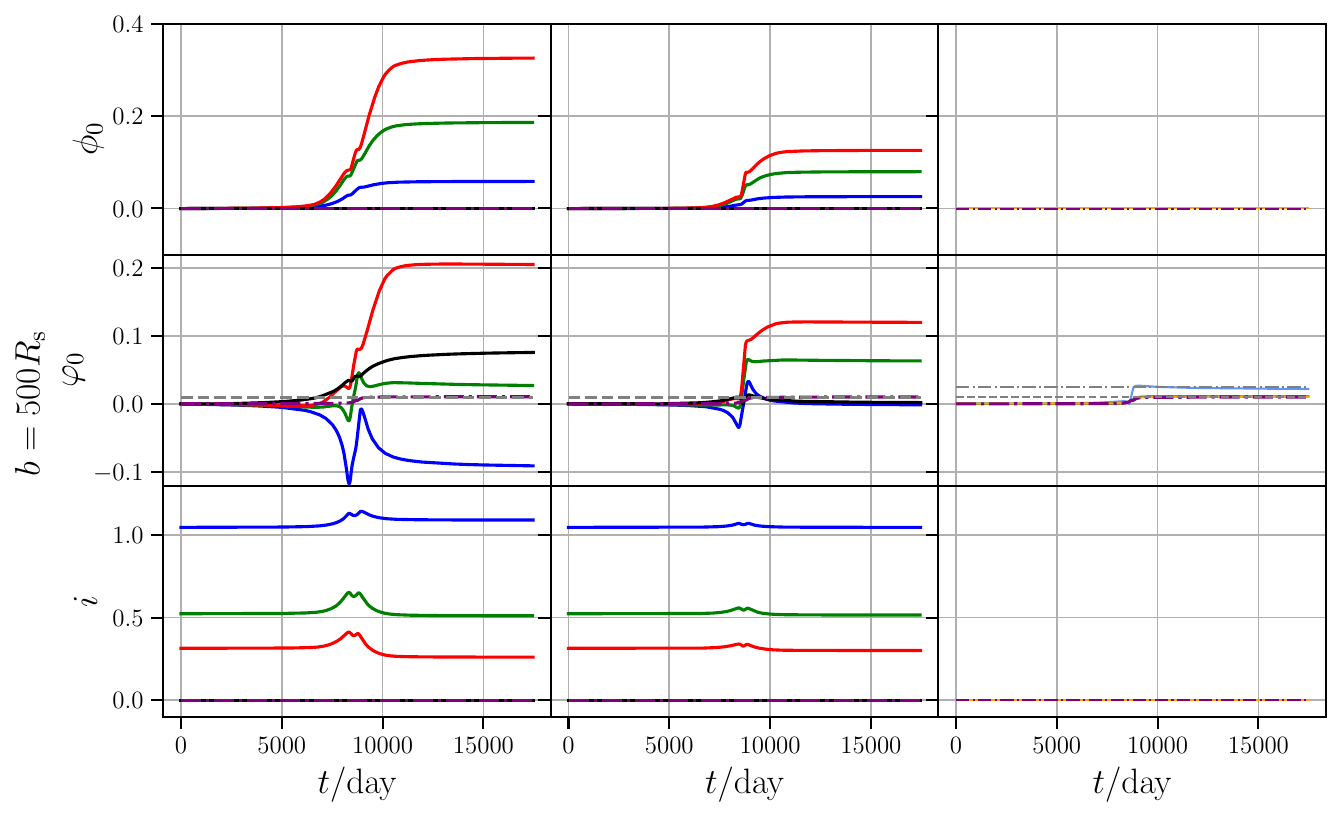}
	\caption{Evolution of angular osculating orbital elements $\{\phi_0,\varphi_0,i\}$ during the hyperbolic encounter, with impact parameter $b=50R_\text{s}$ (upper panel) or $500R_\text{s}$ (lower panel). The setting of other parameters is same with Fig.~\ref{fig:orbit_para1}. The horizontal gray dashed line is the analytical approximation to the total periastron shift due to the 1PN correction given by Eq.~\eqref{1PN_Delta_varphi0}.}
\label{fig:orbit_para2}
\end{figure*}
Using the models of DM spike and accretion disks introduced in the last section, we now examine the orbital evolution of EMRHEs in the vacuum and nonvacuum scenarios. Figure.~\ref{fig:orbit_incli} plots the trajectories of CO projected onto the $x'y'$ and $y'z'$ planes for a given set of parameters. We find that the effect of the accretion disk on the scattering angles is stronger than that of the DM spike, and for EMRHEs in the presence of the accretion disk, deviations between the vacuum and nonvacuum trajectories are more distinctive in the case of inclined orbits.  We also compare the trajectories with impact parameter $b=50R_{\text{s}}$ and $b=500R_{\text{s}}$.  The accretion disk typically results in a larger variation of scattering angles at smaller impact parameter, while at near-zero (but finite) inclination, the situation is reversed.

We reconstruct the evolution of osculating elements using the method described in Appendix~\ref{elements}. Figure.~\ref{fig:orbit_para1} shows the evolution of orbital eccentricity $e$ and semimajor axis $a$. Since the dissipative force (including the DF and GW damping) in all cases is much weaker\footnote{For the gravitational radiation reaction, this is guaranteed by the smallness of $v$ and $q$. For the environmental effects, this is in fact also required by our assumption of the environment being only weakly influenced by the CO; otherwise, the environment cannot be modeled with a stationary density profile.} than the conservative one, and the duration of the encounter is too short to result in any sizable dissipation, the orbital evolution is predominantly driven by the 1PN correction and the environmental gravitational potential. The 1PN effect leads to an increase (decrease) followed by a later decrease (increase) of the semimajor axis (eccentricity), while the environments typically have opposite effects in the initial and final stages of the orbital evolution. Note that the small difference between the initial and final values of the semimajor axis is due to the initial radius being finite.

We also examined the evolution of quantities which should be conserved in the absence of dissipative force. Since the considered environmental gravitational potential $\Phi(\mathbf{r})$ is symmetric about the $z$ axis, the conserved quantities under the acceleration $\mathbf{F}=\mathbf{F}^\text{(1PN)}-\nabla\Phi$ are the orbital energy $\mathcal{E}=\bar E+\Phi(\mathbf{r})$ and $z$-component of the angular momentum $\bar{\mathbf{h}}\cdot\mathbf{e}_z=\bar h\cos i(t)$, with
\begin{align}
\bar E =& E+\frac{3(1-3\nu)}{8}v^4+\frac{M}{2r}\left[(3+\nu)v^2+\nu(\dot r)^2+\frac{M}{r}\right], \nonumber
\\
\bar{\mathbf{h}}=&\left[1+\frac{1-3\nu}{2}v^2+(3+\nu)\frac{M}{r}\right]\mathbf{h}=\bar h\,\mathbf{e}_Z,
\end{align}
where $E=v^2/2-M/r$, $\mathbf{h}=\mathbf{r}\times\mathbf{v}$. Note $\bar E$ and $\bar{\mathbf{h}}$ are the 1PN conserved quantities \cite{maggiore2008gravitational} in the absence of $\Phi$. We find that both $\mathcal{E}$ and $\bar h\cos i$ stay nearly constant (see Fig.~\ref{fig:hz} for a concrete example), confirming that the dissipative effects are negligible.

Figure.~\ref{fig:orbit_para2} shows the evolution of angular osculating elements $\phi_0$, $\varphi_0$ and $i$. The nonspherical gravitational potential of the accretion disk induces a precession of the orbital plane (as measured by the angle $\phi_0$) and the total precession angle decreases with the inclination; this effect is absent in the case of DM spike. As can be seen, the effect of $\alpha$ disk is larger than the $\beta$ disk under similar conditions, and the DM spike has only a tiny influence on the orbital precession, unless both the power-law index $\gamma$ and the impact parameter $b$ are sufficiently large. This hierarchy of effects is roughly due to the difference in the environmental gravitational potential at the outer region, where most of the time in the encounter is spent. The total mass $M_\text{env}=\int d^3r\,\rho(\mathbf{r})$ of the $\alpha$ disk is larger than that of the $\beta$ disk, hence the gravitational potential of the former is steeper at large radius (with $\lim_{r\to\infty}\Phi=-M_\text{env}/r$), as depicted in Fig.~\ref{fig:phi}. The acceleration due to $\beta$ disk can be stronger at the inner region, but whose effect on the orbital evolution turns out to be less significant. The total mass of the considered DM spike does not converge in the limit $r\to \infty$, its enclosed mass in the relevant region is however smaller than that of the accretion disk. Increasing the power-law index of DM spike or the viscosity coefficient of accretion disk leads to an overall increase of density, thereby enhancing their effects.

In the case of DM spike, the density profile \eqref{spike_density} produces a power-law radial acceleration: $\mathbf{F}=-A r^{-n}\,\mathbf{e}_r$ for $r\gg R_\text{s}$ if $\xi<3$, with $A=4 \pi \rho_\text{sp}r_\text{sp}^\xi/(3-\xi)$ and $n=\xi-1$. The total periastron shift can be estimated using Eq.~\eqref{power-law} (since $\xi>2$, $n>1$ is automatically satisfied) together with Eq.~\eqref{1PN_Delta_varphi0}, as indicated by the gray dot-dashed line in Fig.~\ref{fig:orbit_para2} for $\gamma=2$, in agreement with the numerical result.

\begin{figure*}[ht]
	\centering
    \includegraphics[width=0.48\textwidth]{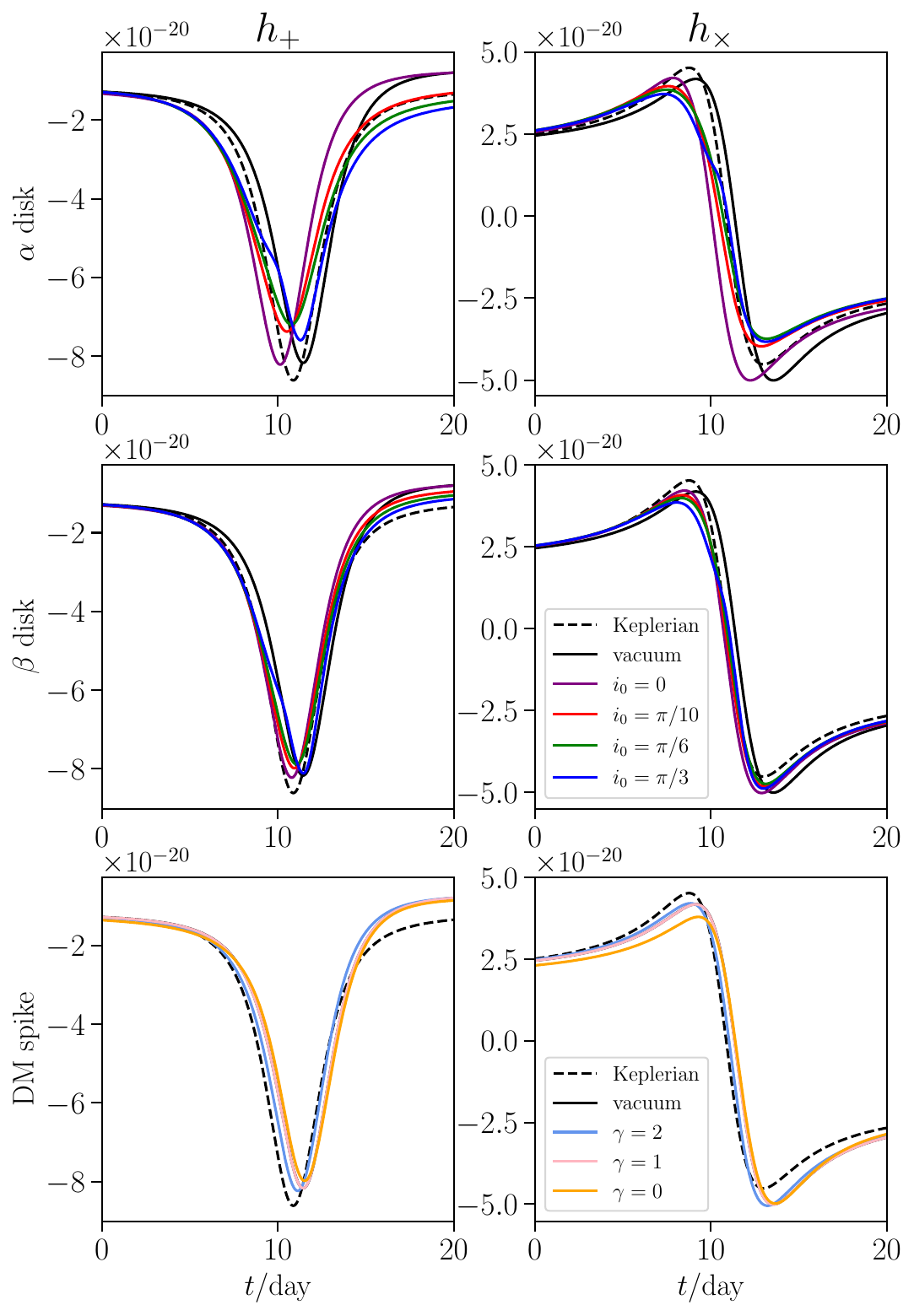}
    \includegraphics[width=0.48\textwidth]{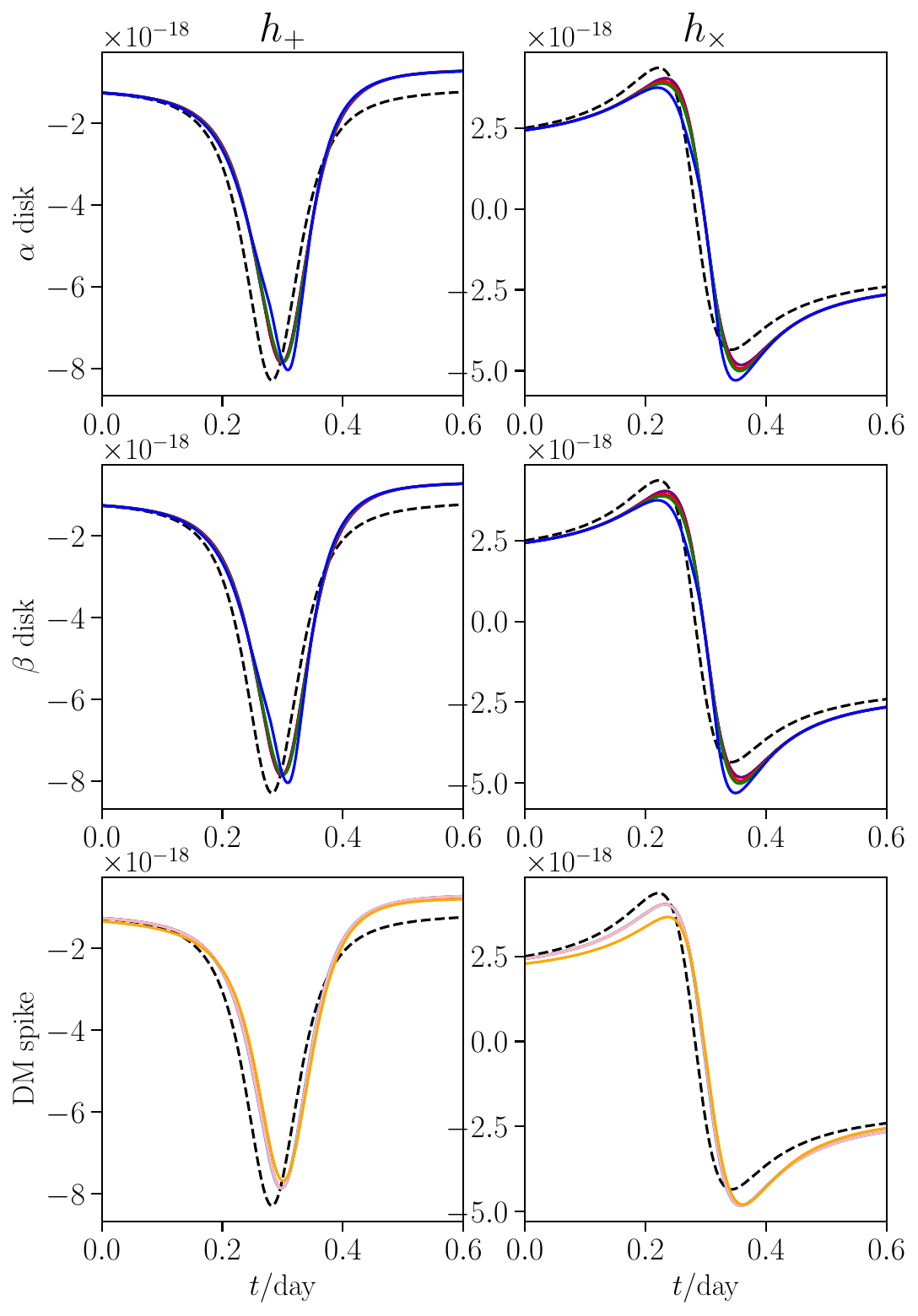}
	\caption{Time-domain waveforms of EMRHEs around a SMBH with mass $m_1=1.4\times 10^8 \, M_{\odot}$ (left two columns) or $4\times 10^6 \, M_{\odot}$ BH (right two columns) in different BH environments, with the impact parameter $b=50 \,R_{\text{s}}$ and initial eccentricity $e_0=2$. The mass of the secondary body is $m_2=10\, M_{\odot}$. The viscosity coefficient of the accretion disk is set to be $\alpha=0.02$. In comparison, the waveforms corresponding to the initial osculating Keplerian orbits are shown in dashed curves.}
\label{fig:GW_wave1}
\end{figure*}

\begin{figure*}[ht]
	\centering
    \includegraphics[width=1.0\textwidth]{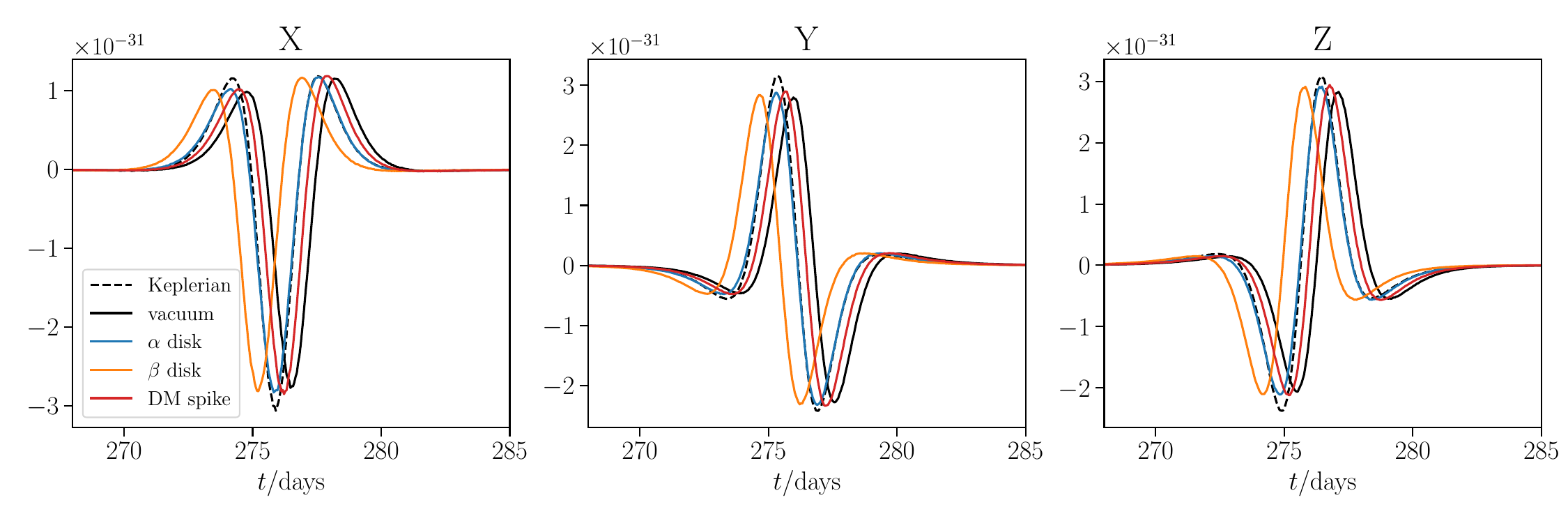}
    \includegraphics[width=1.0\textwidth]{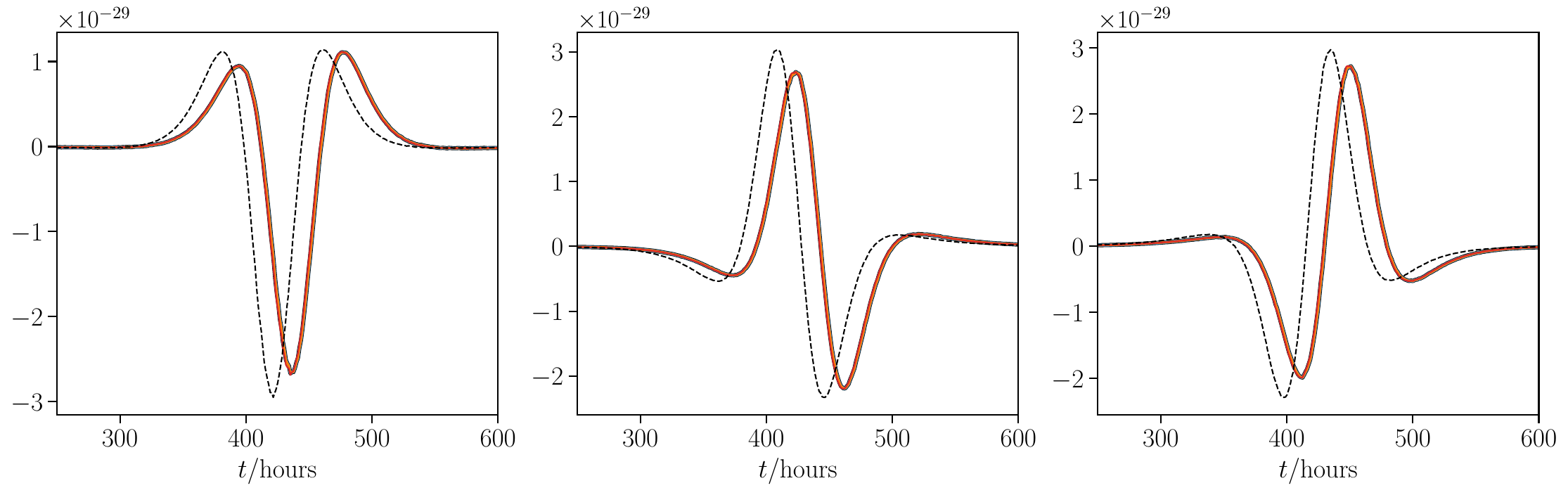}
	\caption{Time-domain outputs of the TDI-2 $X,Y,Z$ combinations to the waveforms with $i_0=0$ and $m_1=1.4\times 10^8 \, M_{\odot}$ (upper panel) or $4\times 10^6 \, M_{\odot}$ (lower panel) in Figure.~\ref{fig:GW_wave1}.}\label{tdi_wave}
\end{figure*}

\begin{figure*}[ht]
	\flushleft
    \includegraphics[width=0.95\textwidth]{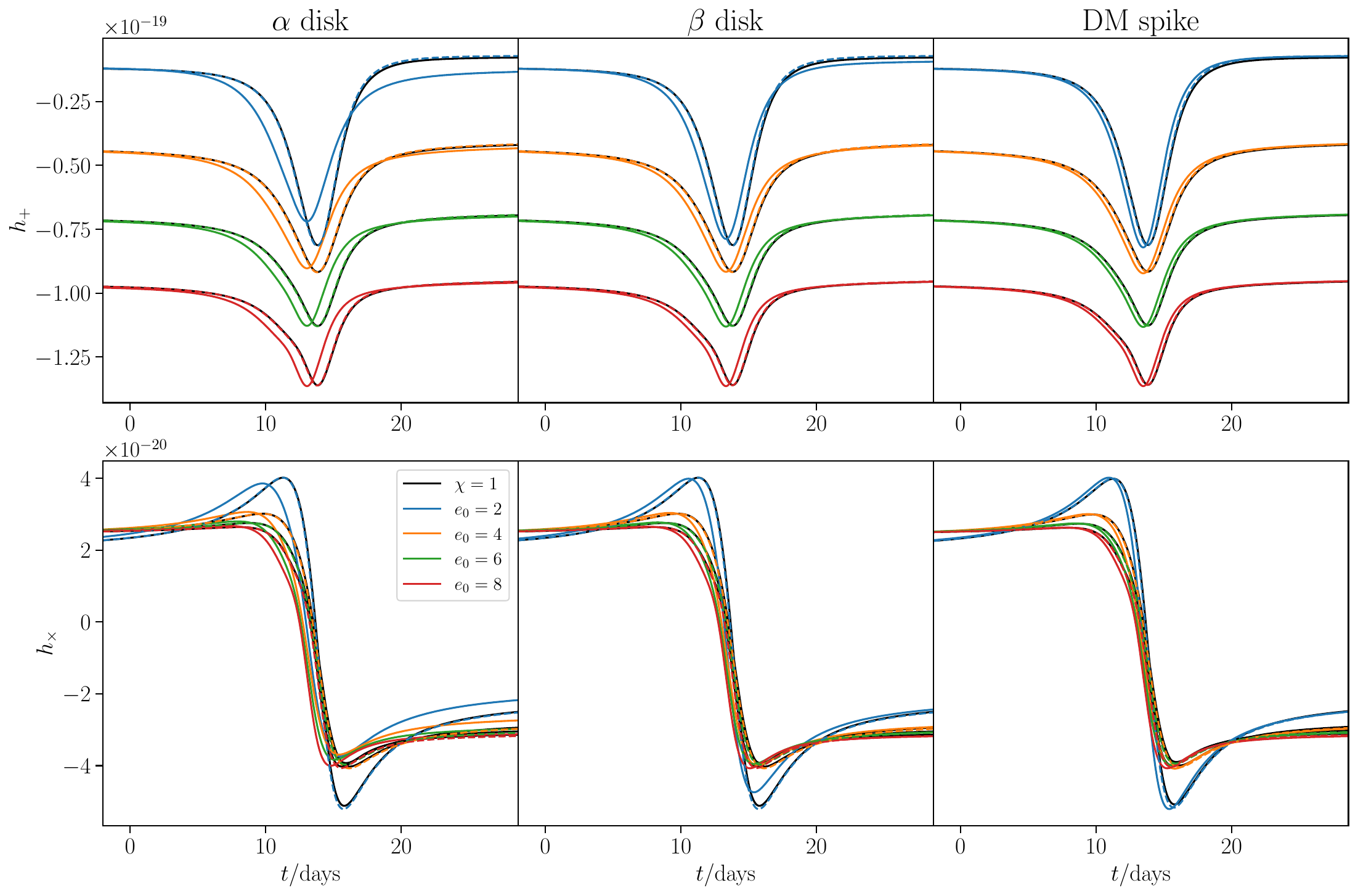}
	\caption{Time-domain waveforms of EMRHEs around M31* with the impact parameter $b=50 R_{\text{s}}$, orbital inclination $i_0=\pi/6$ and different initial eccentricity $e_0$. The viscosity coefficient of the accretion disk is set to be $\alpha=0.02$ and the DM spike power law index is set to be $\gamma = 2$. The solid (dashed) curves correspond to results in the presence (absence) of environments.}
 \label{fig:GW_e}
\end{figure*}

\section{Gravitational Waves}\label{sec5}
The modification on the orbital motion of an EMRHE can in principle leave observable imprints in its gravitational wave signals. In this section, we examine the GW waveforms of EMRHEs in the presence of different BH environments and assess their detectabilities by the LISA detector.

To compute the GW waveforms, we use the leading-order quadrupole formula \cite{maggiore2008gravitational}
\begin{equation}\label{eq1}
\begin{aligned}
\left[h_{ij}^\text{TT}(t)\right]_\text{0PN}=\frac{2}{D}\Lambda_{ij,kl}(\mathbf{n})\,\ddot{M}_{kl}(t-D),
\end{aligned}
\end{equation}
where $\mathbf{n}$ is the unit vector pointing from the source to the detector, $D$ is the distance to the detector, and $\Lambda_{ij,kl}(\mathbf{n})$ is the transverse-traceless (TT) projector. The quadrupole moment in the binary's c.m. frame is $M_{ij}=\mu\,r_{i}r_{j}$. For $\mathbf{n}=(\theta_\text{s},\phi_\text{s})$, the strain tensor can be decomposed into the $+$ and $\times$ modes as
\begin{equation}
h_{ij}^\text{TT}(t)=e^+_{ij}\,h_+(t)+e^\times_{ij}\,h_\times(t),
\end{equation}
with the polarization tensors given by
\begin{align}
e^+_{ij}&=(\mathbf{e}_\theta)_i(\mathbf{e}_\theta)_j-(\mathbf{e}_\phi)_i(\mathbf{e}_\phi)_j,
\\
e^\times_{ij}&=(\mathbf{e}_\theta)_i(\mathbf{e}_\phi)_j+(\mathbf{e}_\theta)_j(\mathbf{e}_\phi)_i.
\end{align}
The explicit waveforms of the two polarization modes are
\begin{align}
(h_{\times})_\text{0PN}=&\frac{1}{D}\big[(\ddot{M}_{22}-\ddot{M}_{11})\,s_{2\phi_\text{s}}c_{\theta_\text{s}}+2\ddot{M}_{12}\,c_{2\phi_\text{s}}c_{\theta_\text{s}} \nonumber\\
&\quad+2\ddot{M}_{13}\,s_{\phi_\text{s}}s_{\theta_\text{s}}-2\ddot{M}_{23}\,c_{\phi_\text{s}}s_{\theta_\text{s}}\big],
\\
    (h_{+})_\text{0PN}=&\frac{1}{D}
    \big[\ddot{M}_{11}\,(-s^2_{\phi_\text{s}}+c^2_{\phi_\text{s}}c^2_{\theta_\text{s}})
    -\ddot{M}_{23}\,s_{\phi_\text{s}}s_{2\theta_\text{s}}\nonumber\\
    &\quad+\ddot{M}_{22}\,(-c^2_{\phi_\text{s}}+s^2_{\phi_\text{s}}c^2_{\theta_\text{s}})+\ddot{M}_{33}\,s^2_{\theta_\text{s}}\nonumber\\
    &\quad+\ddot{M}_{12}\,s_{2\phi_\text{s}}(1+c^2_{\theta_\text{s}})-\ddot{M}_{13}\,c_{\phi_\text{s}}s_{2\theta_\text{s}}
    \big],
\end{align}
where $c_z=\cos z$ and $s_z=\sin z$. Note that these differ from the formulas in \cite{maggiore2008gravitational} since we have used the standard definition of spherical coordinates $(\theta,\phi)$.

To improve the accuracy, we also include the 0.5PN correction to the waveform (see for example \cite{Roskill:2023bmd}),
\begin{equation}
\left[h_{ij}^\text{TT}(t)\right]_\text{0.5PN}=\frac{1}{D}\Lambda_{ij,kl}(\mathbf{n})\,n_m\left(\frac{2}{3}\dddot M_{klm}+\frac{4}{3}\ddot{\mathcal{Z}}_{klm}\right),
\end{equation}
where the mass octupole moment $M_{klm}$ and the current quadrupole moment $\mathcal{Z}_{klm}$ can be evaluated for $m_2<m_1$ as
\begin{align}
M_{klm}&=\mu \sqrt{1-4\nu}\,r_kr_lr_m,
\\
\mathcal{Z}_{klm}&=P_{k,lm}+P_{l,km}-2P_{m,kl},
\end{align}
with $P_{k,lm}=\mu \sqrt{1-4\nu}\,\dot r_k r_lr_m$. In all cases we considered, $(h_{+,\times})_\text{0.5PN}$ are already small, so the higher PN corrections \cite{Bini:2021jmj,Cho:2022pqy} (e.g., that from the 1PN correction to $M_{ij}$) will be neglected.

For the space-based interferometers, time-delay interferometry (TDI) is generally required in data analysis. Here we consider the second-generation TDI (TDI-2) observables \cite{Tinto:2004wu,babak2021lisasensitivitysnrcalculations,QuangNam:2022gjz}, adopting for simplicity the equal-arm approximation with arm length $L=2.5\,\text{Gm}$ for the LISA detector. For convenience, we choose the $x'y'z'$ frame to be a translation of the solar-system-barycenter (SSB) frame used by \cite{Krolak:2004xp} (that is, we consider the case of BH spin perpendicular to the ecliptic plane), such that the source direction is given by $(\theta_\text{S},\phi_\text{S})=(\pi-\theta_\text{s},\pi+\phi_\text{s})=(-\beta+\pi/2,\lambda)$ with $\psi=\pi/2$. For concreteness, we also set $(\theta_\text{s},\phi_\text{s})=(i_0,\phi_0+\pi)$, so that $\mathbf{n}$ is normal to the initial orbital plane and the waveforms for different initial inclination $i_0$ are same in the vacuum case. For the orbit configuration of LISA, we use the setting of \cite{Krolak:2004xp}, with $\zeta=-\pi/6, \Omega=2\pi/\text{yr}$ and $\xi_0=\eta_0=0$.

\begin{figure*}[ht]
	\centering
    \includegraphics[width=0.497\textwidth]{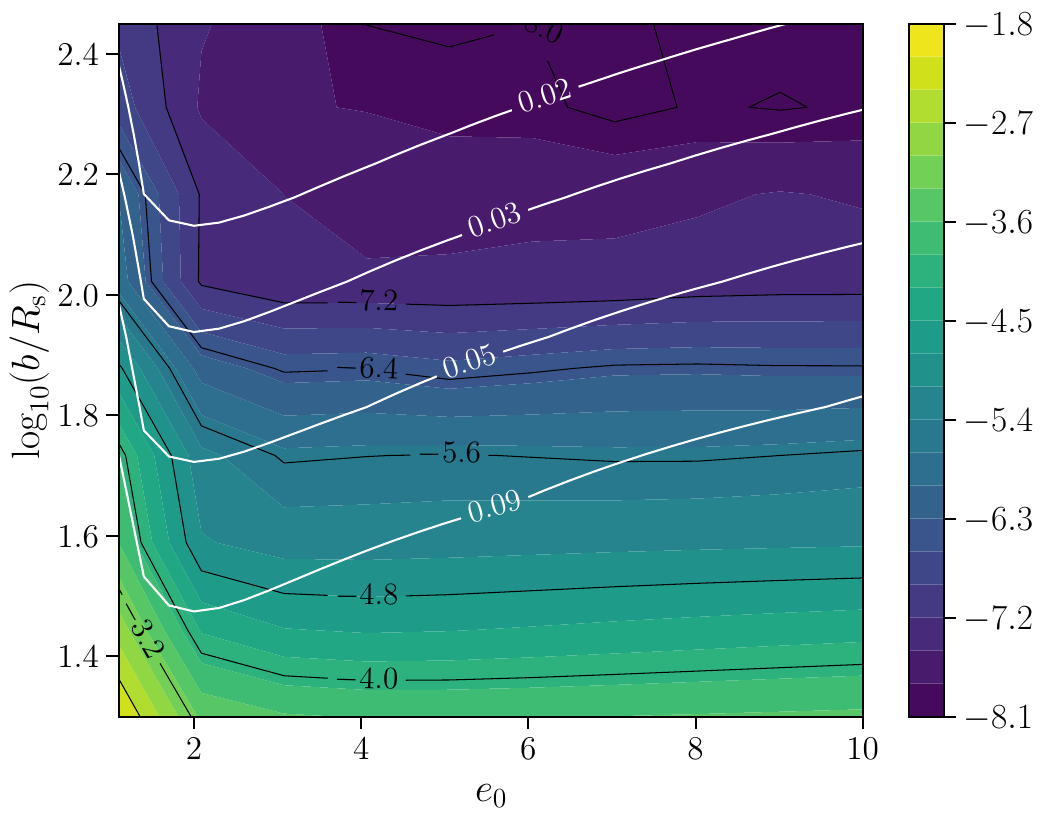}
    \includegraphics[width=0.495\textwidth]{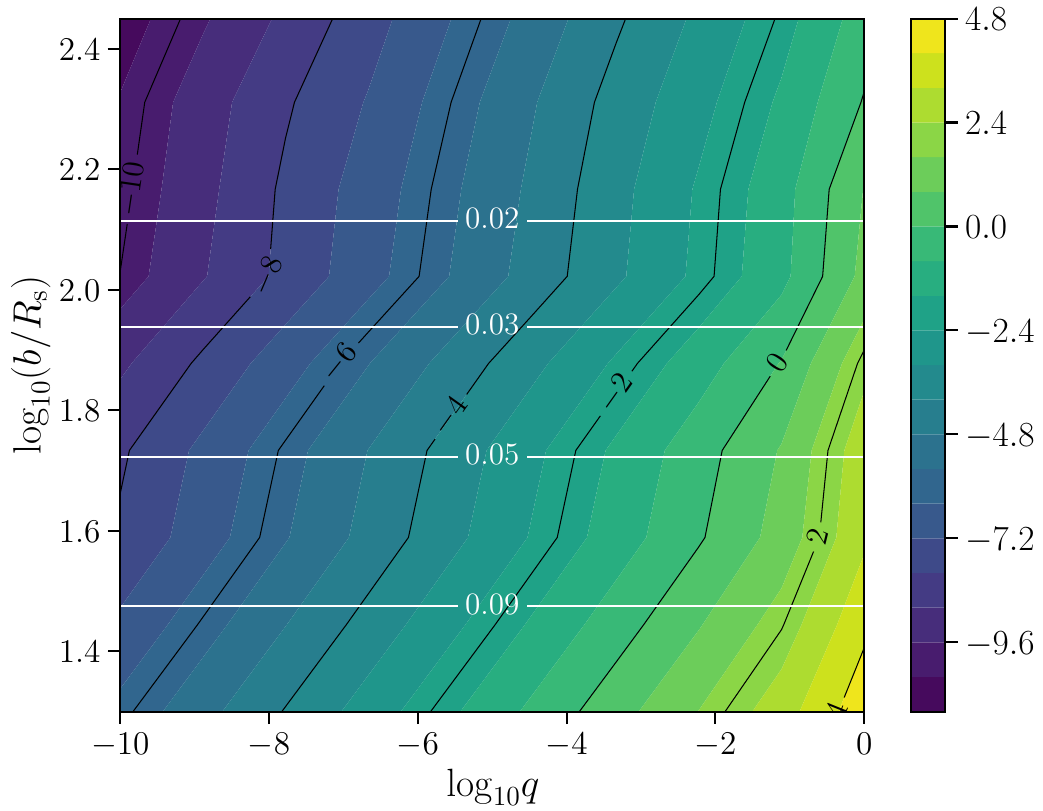}
\\
    \includegraphics[width=0.497\textwidth]{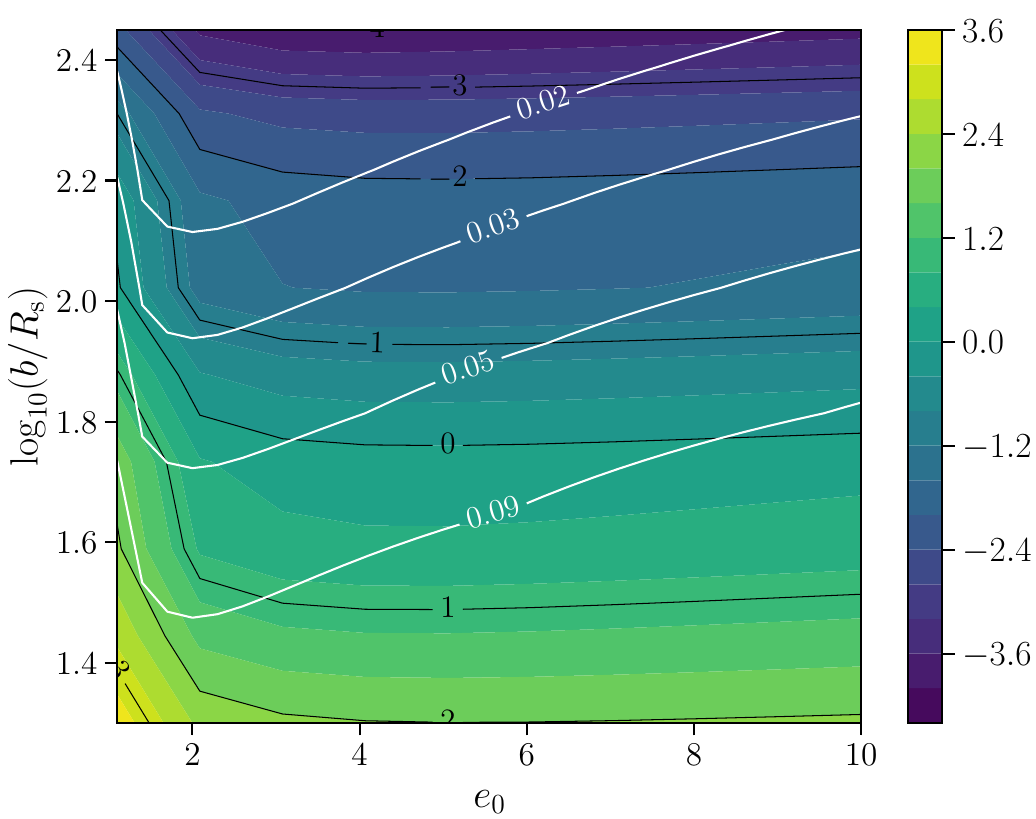}
    \includegraphics[width=0.495\textwidth]{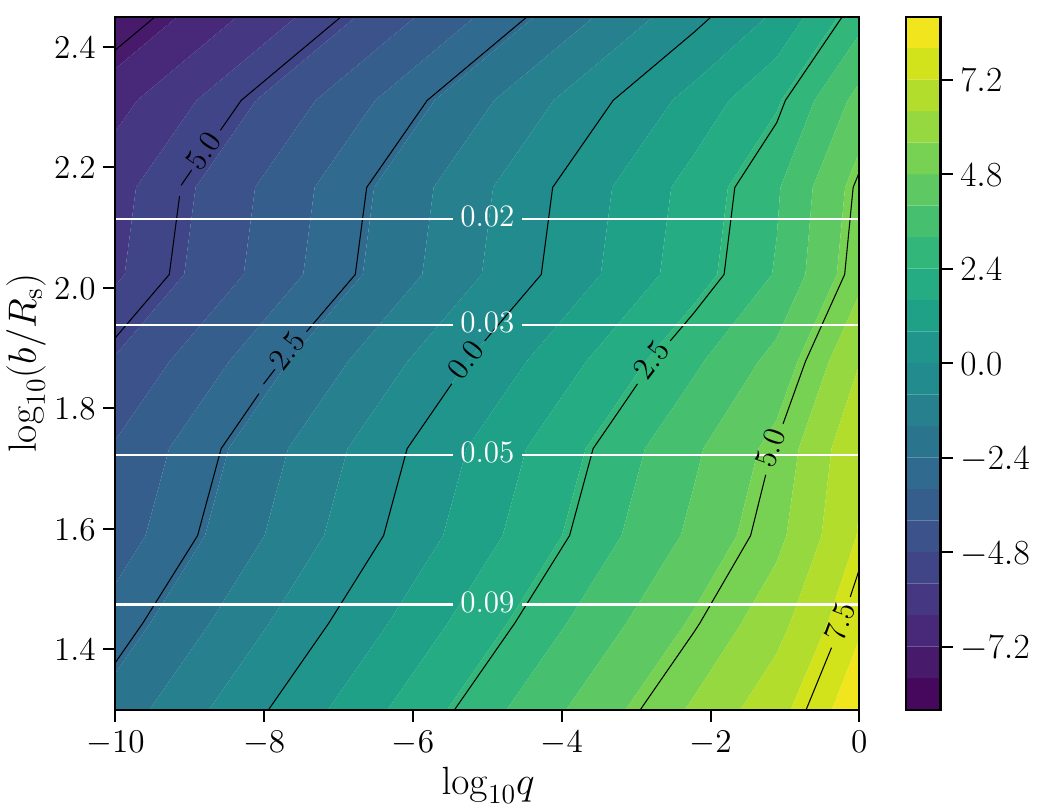}
    \caption{$\log_{10}\text{SNR}_\text{X}$ of EMRHEs around M31* (upper row) and Sgr A* (lower row) at LISA as functions of initial eccentricity $e_0$ (left column) or mass ratio $q=m_2/m_1$ (right column) and the impact parameter $b$, with given CO mass $m_2=10 \, M\odot$ (left column) or initial eccentricity $e_0=2$ (right column). We set $i_0=0$, and the contours of $v_\text{max}^2\approx [(e_0+1)/(e_0-1)](M/a_0)$ are shown in white curves.}
 \label{fig:snr}
\end{figure*}

We consider two sets of EMRHE events around \footnote{
For Sgr A*, the standard disk model is unfavorable in view of the constraints from electromagnetic observations \cite{Hosseini_2020}, The environmental effects due to accretion disk around Sgr A* is thus expected to be weaker than our considered models.
}Sgr A* \cite{2012Sci...338...84M} (with $m_1\approx 4\times \, 10^6 M_\odot$, $D\approx8000\, \text{pc}$) and M31* \cite{2005ApJ...631..280B,Karachentsev:2004dx} (with $m_1\approx 1.4\times 10^8 \, M_\odot$, $D\approx0.77\,\text{Mpc}$). The GW waveforms $h_{+,\times}(t)$ are shown in Fig.~\ref{fig:GW_wave1}. The modifications to the orbital evolution are reflected in the time-domain waveforms, including the variations in the peak position and the amount of displacement memory $\Delta h=h(\infty)-h(-\infty)$. The corresponding TDI signals $\{X(t), Y(t), Z(t)\}$ are shown in Fig.~\ref{tdi_wave}.

In the case of the accretion disk, the peak shift is most pronounced when the inclination angle is small, in which case the orbit remains close to the disk plane for a relatively long time and the orbital evolution is affected by the gravitational potential of the disk at an early stage. As the orbital inclination decreases, the waveform peak of $+$ mode becomes steeper, accompanied by an enhancement in its GW memory. The difference between vacuum and nonvacuum waveforms is most visible in the case of $\alpha$ disk; in the case of the DM spike, the difference increases with the power-law index $\gamma$. Given the same CO mass, impact parameter and initial eccentricity, the vacuum and nonvacuum waveforms are harder to distinguish for events around Sgr A* compared to those around M31*. 

The time-domain waveforms of EMRHE events with varying initial eccentricities are shown in Fig.~\ref{fig:GW_e}. We find that decreasing the orbital eccentricity generally increase the peak amplitude as well as the difference between vacuum and nonvacuum waveforms. As a check for our omission of the spin effects, we also show the vacuum waveform (black curve in Fig.~\ref{fig:GW_e}) after including the leading-order spin-orbit coupling \eqref{spin} in the orbital dynamics for a maximally spinning central BH, whose deviation from the case of zero spin is indeed small.

Finally we examine the signal-to-noise ratio of EMRHE events based on the TDI-2 X observable. For a signal $X(t)$, the optimal SNR in a matched filtering is given by
\begin{equation}
\text{SNR}_\text{X}^2=
\int d \ln f\,\left(\frac{|2f\tilde{X}|}{\sqrt{f P_n}}\right)^2
,
\end{equation}
where $\tilde{X}(f)=\int dt\,e^{2\pi ift}X(t)$ and $P_n(f)$ is the one-sided power spectral density (PSD) of the detector noise \cite{babak2021lisasensitivitysnrcalculations}. The $\text{SNR}_\text{X}$ for EMRHE events around Sgr A* and M31* in the vacuum case are shown in Fig.~\ref{fig:snr}. For given initial eccentricity, the square of maximum velocity reached during the encounter can be estimated by the Keplerian value $v_\text{max}^2\approx [(e_0+1)/(e_0-1)](M/a_0)$ (as indicated by the red lines), which remains small throughout the depicted parameter space, justifying the use of 1PN approximation. For each event, the sampling time range of the waveform corresponds to the orbital evolution from $\varphi=\varphi(t=0)$ to $\varphi\approx-\varphi(t=0)$. In the case of Sgr A* source, the $\text{SNR}_\text{X}$ at LISA reaches $10^1\textendash10^3$ when the impact parameter is about several tens of Schwarzschild radii, such events are likely detectable by the the space-based GW detectors. A similar event near M31* can only have a $\text{SNR}_\text{X}$ around $10^{-4}\textendash10^{-2}$. As concrete examples, we plot the signal spectrum for two sets of EMRHE events in Fig.~\ref{fig:GW_wave3}. The most detectable part of the signal appears to be the spectral peak, whose position can be roughly estimated by the frequency $f_*=\sqrt{(e_0+1)/(e_0-1)^3}\,\sqrt{M/a_0^3}$, corresponding to the angular velocity at periastron for the initial osculating Keplerian orbit.

\section{Conclusion and Discussion}\label{sec6}
We have investigated the orbital evolution and GW waveforms of EMRHEs near a supermassive black hole surrounded by an accretion disk or a DM spike. Our computation incorporates the conservative and dissipative environmental effects as well as the leading PN corrections. For each type of environment, the evolution of osculating orbital elements is tracked and analyzed. In the case of accretion disk, we construct a three-dimensional model for its gravitational potential, which allows us to examine the full range of initial conditions for the hyperbolic scattering. We find that the primary influence of a black hole environment on the hyperbolic encounter comes from its gravitational potential rather than the dissipative effects such as dynamical friction. Additionally, the impact of DM spike on the orbital evolution is found to be typically smaller than that of the accretion disk in our considered models, especially for an orbit with finite inclination.

We also demonstrate the detectability of such extreme mass-ratio encounters and the discriminability of BH environments in the waveforms. We find that events which show clear modifications of waveforms due to the presence of environments are typically below the sensitivity of near future LISA-like detectors. On the other hand, for detectable events the imprints of environments on GW waveforms are small, thus requiring more sensitive detectors or more advanced analysis methods to distinguish them.

As possible extensions of the present work, other types of accretion disks can be investigated, such as the slim disk model \cite{1989ASSL..156..273L} describing super-Eddington accretion and the advection dominated accretion flow model \cite{1998tbha.conf..148N} describing hot accretion. Other types of BH environments may also be considered, such as the nonrelativistic cloud of ultralight bosons\footnote{See \cite{Tomaselli:2023ysb} for an estimation of the energy dissipation of hyperbolic orbit in the scalar $|211\rangle$ cloud.}, for which the environmental gravitational potential is fixed by the wave function of the cloud~\cite{Cao:2023fyv}. Although the dissipative effects vary in different types of environments \cite{1989ApJ...336..313P,Barausse:2007ph,Vicente:2022ivh,Traykova:2023qyv} and are subject to relatively large theoretical uncertainty, for the fast hyperbolic encounter it is expected that the conservative gravitational effect is generally much more important, as long as the environment is only weakly perturbed. The dissipative effects would become important only for sufficiently small incident velocity, as in the extreme case of dynamical capture. Besides the environmental effects, modifications to the intrinsic binary dynamics (e.g., due to the presence of scalar or vector charges~\cite{PhysRevD.103.023015, Cheng:2023qys}) can also be examined. Going beyond the Newtonian regime, a fully relativistic description of the binary as well as the environments~\cite{Speeney:2024mas} is needed for an accurate modeling of such systems.

\section*{acknowledgement}
This work is supported by the National Key Research and Development Program of China (No. 2021YFC2201901), the National Natural Science Foundation of China (No.~12347103) and the Fundamental Research Funds for the Central Universities.
\newpage
\appendix
\section{Initial condition and reconstruction of osculating elements}\label{elements}
\begin{table}[t]
\renewcommand{\arraystretch}{1}
    \begin{center}
        \begin{tabular}{ccccccc}
        \hline
        \hline
         \quad Central BH \quad & $\quad m_1 \quad$ & $\quad \rho_0 /\text{g}\,\text{cm}^{-3} \quad$ & $\qquad r_0 /\text{pc} \qquad$\\
         \hline
         Sgr A* &$4\times 10^6 M_\odot$& $7.7\times10^{-25}$& $1.8\times 10^3$\\
          M31*  &$1.4\times 10^8 M_\odot$& $1.98\times 10^{-25}$&$3.46\times 10^4$\\ 
         \hline
         \hline
        \end{tabular}
    \end{center}
    \caption{Estimated parameters of the initial DM halos for the Milky Way and the M31 galaxy.}
    \label{table0}
\end{table}
Given the initial osculating elements $\{a,e,\varphi_0,\phi_0,i\}$ and true anomaly $\varphi$, the intial orbital radius is
\begin{equation}
r=\frac{a(e^2-1)}{1+e\cos{\varphi}},
\end{equation}
with the Cartesian components of position $\mathbf{r}=x'\mathbf{e}_{x'}+y'\mathbf{e}_{y'}+z'\mathbf{e}_{z'}$ and velocity $\mathbf{v}=v_{x'}\mathbf{e}_{x'}+v_{y'}\mathbf{e}_{y'}+v_{z'}\mathbf{e}_{z'}$ in the fixed $x'y'z'$ coordinate system (see Fig.~\ref{fig:coor}) given by
\begin{align}
\frac{x'}{r}&=\cos{i}\,\cos{\phi_0}\cos{(\varphi+\varphi_0)}-\sin{\phi}\sin{(\varphi+\varphi_0)},
\nonumber\\
\frac{y'}{r}&=\cos{i}\,\sin{\phi_0}\cos{(\varphi+\varphi_0)}+\cos{\phi_0}\sin{(\varphi+\varphi_0)},
\nonumber\\
\frac{z'}{r}&=\sin{i}\,\cos{(\varphi+\varphi_0)},
\end{align}
and
\begin{align}
    v_{x'}&=\sqrt{\frac{M}{a(e^2-1)}}\bigg\{-\sin{\phi_0}[e\cos{\varphi_0}+\cos{(\varphi+\varphi_0)}]\nonumber
    \\
    &\quad -\cos{i}\,\cos{\phi_0}\,[e\sin{\varphi_0}+\sin{(\varphi+\varphi_0)}]\bigg\},
\nonumber\\
    v_{y'}&=\sqrt{\frac{M}{a(e^2-1)}}\bigg\{\cos{\phi_0}\,[e\cos{\varphi_0}+\cos{(\varphi+\varphi_0)}]\nonumber
    \\
    &\quad -\cos{i}\,\sin{\phi_0}\,[e\sin{\varphi_0}+\sin{(\varphi+\varphi_0)}]\bigg\},
\nonumber\\
    v_{z'}&=-\sqrt{\frac{M}{a(e^2-1)}}\sin{i}\,[e\sin{\varphi_0}+\sin{(\varphi+\varphi_0)}].\label{velocity}
\end{align}

Once a trajectory $\mathbf{r}(t)$ is computed, the osculating orbital elements can be obtained as follows \cite{Curtis2020}. Introducing the vectors of angular momentum, ascending node and eccentricity:
\begin{align}
\mathbf{L} & =\mathbf{r}\times\mathbf{v},
\\
\mathbf{N} & =\mathbf{e}_{z'}\times\mathbf{L},
\\
\mathbf{e} & =\frac{1}{M}\left[\left(v^2-\frac{M}{r}\right)\mathbf{r}-(\mathbf{r}\cdot\mathbf{v})\,\mathbf{v}\right],
\end{align}
we have
\begin{align}
a &= \left(\frac{2}{r}-\frac{v^2}{M}\right)^{-1},
\\
e&=|\mathbf{e}|,
\\
i & =\arccos\left(\frac{\mathbf{L}\cdot\mathbf{e}_{z'}}{|\mathbf{L}|}\right),
\\
\varphi_0 & =
\begin{cases}
\arccos\left(\frac{\mathbf{N}\cdot\mathbf{e}}{|\mathbf{N}|e}\right)-\frac{\pi}{2},&\mathbf{e}\cdot\mathbf{e}_{z'}\geq 0
\\
-\arccos\left(\frac{\mathbf{N}\cdot\mathbf{e}}{|\mathbf{N}|e}\right)+\frac{3\pi}{2},& \mathbf{e}\cdot\mathbf{e}_{z'}<0
\end{cases}
\\
\phi_0 & =
\begin{cases}
    \arccos\left(\frac{\mathbf{N}\cdot\mathbf{e}_{x'}}{|\mathbf{N}|}\right)-\frac{\pi}{2}, &\mathbf{N}\cdot\mathbf{e}_{y'}\geq0
    \\
    -\arccos\left(\frac{\mathbf{N}\cdot\mathbf{e}_{x'}}{|\mathbf{N}|}\right)+\frac{3\pi}{2}, &\mathbf{N}\cdot\mathbf{e}_{y'}<0
\end{cases}
\\
\varphi &=
\begin{cases}
    \arccos\left(\frac{\mathbf{e}\cdot\mathbf{e}_r}{e}\right), & \dot r \geq 0
    \\
    -\arccos\left(\frac{\mathbf{e}\cdot\mathbf{e}_r}{e}\right)+2\pi,& \dot r < 0
\end{cases}
\end{align}
\section{Density distribution of DM spike}\label{app_dm}
We follow \cite{Gondolo:1999ef,Sanchez-Conde:2013yxa,Nishikawa:2017chy} for the modeling of DM spike. The parameters $(\rho_\text{sp},r_\text{sp})$ in Eq.~\eqref{spike_density} for given power-law index $\gamma$ are determined from
\begin{equation}
    \rho_{\text{sp}}=\rho_0 {(r_{\text{sp}}/r_0)}^{-\gamma},
\quad
    r_{\text{sp}}=\alpha_\gamma r_0 {(m_1/\rho_0 r_0^3)}^{1/(3-\gamma)},
\end{equation}
where $\alpha_\gamma$ is a normalization factor, $(\rho_0,r_0)$ are the parameters of the initial DM halo. Assuming the initial halo follows the Navarro-Frenk-White profile, $(\rho_0,r_0)$ can be obtained by solving the following set of equations:
\begin{align}
\sigma^{2}&=4\pi \rho_0\,r_0^2\,c_\text{m}^{-1}\,g(c_\text{m}),
\\
 M_{\text{vir}}&=4 \pi \rho_0 \,r_0^3 \,g(R_{\text{vir}}/r_0)=
 \frac{4}{3} R_{\text{vir}}^3\,\rho_\text{c}\, \Delta,
\end{align}
where $c_{\text{m}}=2.16$, $g(x)=\ln{(1-x)}-x/(1+x)$, $\Delta \approx 200$, $\rho_\text{c}$ is the cosmic critical density, $M_{\text{vir}}$ is the virial mass, $R_{\text{vir}}$ is the virial radius, $\sigma$ is the one-dimensional halo velocity dispersion given by
\begin{equation}
    \log_{10}(m_1/M_\odot)=a+b \log_{10}\left(\sigma/200\,\text{km s}^{-1}\right),
\end{equation}
where $a=8.12$, $b=4.24$ and $R_{\text{vir}}/r_0=\sum_{i=0}^5c_i\,z^i$, with $z=\ln \left(h_0 M_{\text{vir}}/M_\odot\right)$, $h_0=0.67$, $c_0=37.5153$, $c_1=-1.5093$, $c_2=1.636 \times 10^{-2}$, $c_3=3.66\times 10^{-4}$, $c_4=-2.89237\times 10^{-5}$, $c_5=5.32\times 10^{-7}$ \cite{Sanchez-Conde:2013yxa}. We report the estimated values of $(\rho_0,r_0)$ for Sgr A* and M31* in Table~\ref{table0}.

\begin{figure*}[ht]
	\centering
	\includegraphics[width=0.75\textwidth]{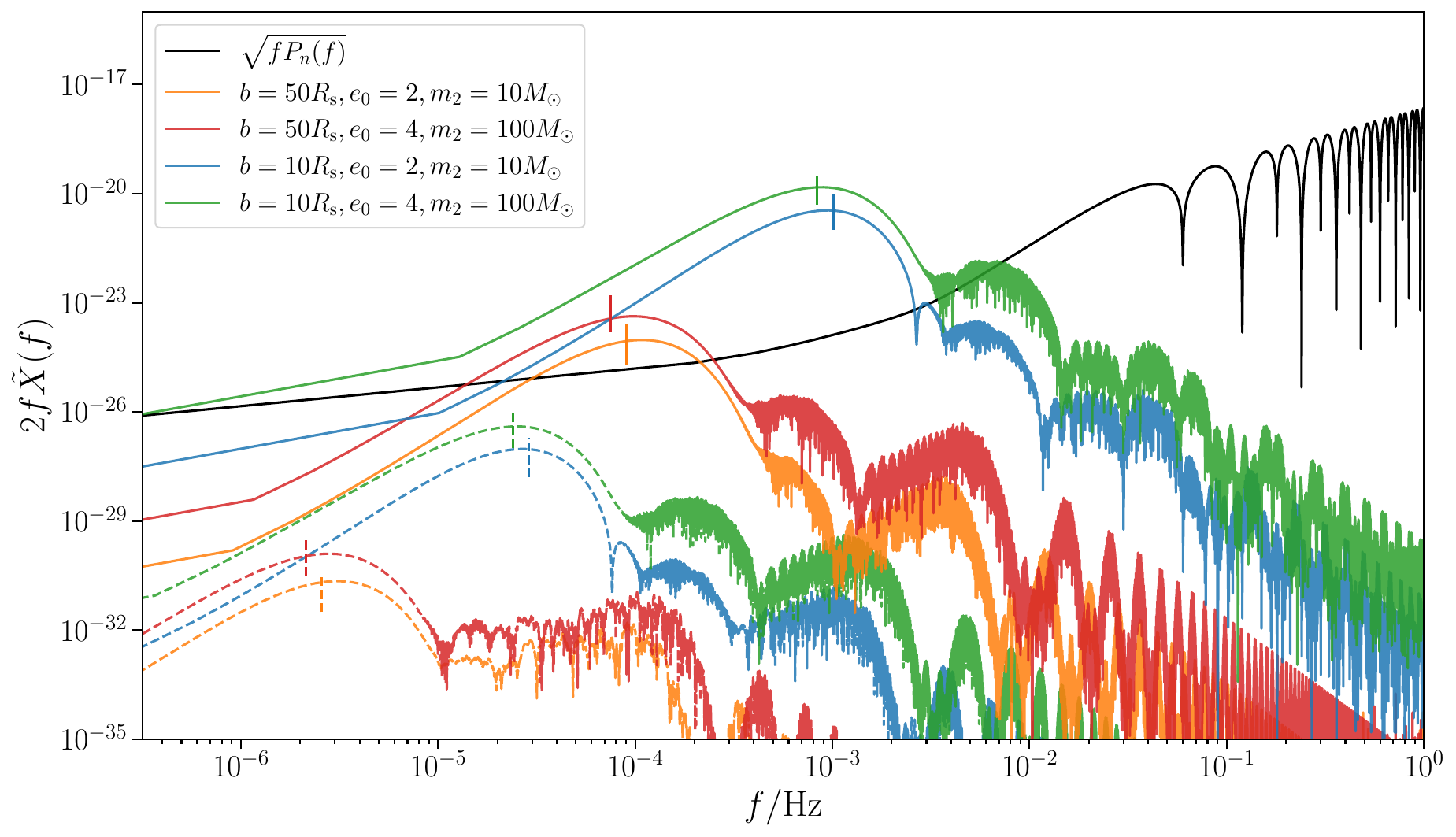}
	\caption{Signal spectrum of the TDI-2 X combination for EMRHE events around Sgr A* (upper solid curves) and M31* (lower dashed curves) in the vacuum case, shown against the noise curve (black solid line). We set $i_0=0$. The frequency $f_*=\sqrt{(e_0+1)/(e_0-1)^3}\,\Omega$ is indicated by the vertical line.}
\label{fig:GW_wave3}
\end{figure*}

\section{Density distribution of accretion disk}\label{disk}

\begin{table}[t]
\renewcommand{\arraystretch}{1}
    \begin{center}
        \begin{tabular}{cccccccc}
        \hline
        \hline
         $\quad$Disk type$\quad$ & $ b $ & $\quad \epsilon \quad$ & $\quad \kappa \quad$ & $\quad \mu \quad$ & $\quad X_\text{H} \quad$ & $\quad l_\text{E} \quad$\\
         \hline
         $\alpha$ &0& $0.1$&$0.349\,\text{cm}^2\text{g}^{-1}$&1.756& 0.75 & 0.1\\
         $\beta$ &1& $0.1$&$0.349\,\text{cm}^2\text{g}^{-1}$&1.756& 0.75 & 0.1\\ 
         \hline
         \hline
        \end{tabular}
    \end{center}
    \caption{The choices for various parameters in the disk model. $\epsilon$ is the radiative efficiency of the disk and $\kappa$ is the gas opacity.}
    \label{table1}
\end{table}

The surface mass density distribution of a standard thin accretion disk is given by  \cite{2004ApJ...608..108G}
\begin{align}\label{Sigma}
    \Sigma_0=&2.56\times10^5(\alpha_{0.3}\,\beta^{b-1})^{-4/5}{l_\text{E}}^{3/5}\epsilon_{0.1}^{-3/5}\hat{\kappa}^{-1/5}\mu^{4/5} \nonumber \\
    &\times\left(\frac{m_1}{10^8M_\odot}\right)^{1/5}\left(\frac{R}{R_\text{Q}}\right)^{-3/5}\text{g}\,\text{cm}^{-2},
\end{align}
where $\hat{\kappa}$ is the opacity (in the unit of electron scattering opacity of the ionized hydrogen, $0.4\,\text{cm}^2\text{g}^{-1}$), $\mu$ the average molecular mass (in the unit of atomic mass of hydrogen), $\alpha_{0.3}=\alpha/0.3$ the Shakura-Sunyaev viscosity coefficient (with \cite{King:2007cu} $\alpha \sim 0.01-0.1$), $\beta=p_\text{gas}/p$ and $\epsilon=L/\dot{M}c^2=0.1\,\epsilon_{0.1}$ the radiative efficiency of the disk. The parameter $b$ determines whether the viscosity depends on the total pressure (called $\beta$ disk) or only the gas pressure (called $\alpha$ disk).

For the parameter $\beta$, we use \cite{2004ApJ...608..108G}
\begin{align}\label{beta}
    \frac{\beta^{\frac{b+4}{10}}}{1-\beta} \approx &\,0.311\,\alpha_{0.3}^{-1/10}\hat{\kappa}^{-9/10}\mu^{-2/5}(\epsilon_{0.1}/l_\text{E})^{4/5}\nonumber\\
    &\times\left(\frac{m_1}{10^8M_\odot}\right)^{-0.1}\left(\frac{R}{R_\text{Q}}\right)^{21/20},
\end{align}
where $l_\text{E}=\dot{m_1}/\dot{m}_\text{Edd}$ is the ratio of the central BH's mass accretion rate to the Eddington rate, typically \cite{Kollmeier:2005cw} $l_\text{E}\sim 0.1-1$. For simplicity, we fix the disk parameters according to Table~\ref{table1}, leaving only a single free parameter $\alpha$.

Within the gravitational stability radius $R_\text{Q}$ of the disk ($r<R_\text{Q}$), radiative pressure dominates, Eq.~\eqref{beta} becomes
\begin{align}
    \beta^{\frac{b+4}{10}}\approx &\,0.281\,\alpha_{0.3}^{-1/10}{l_\text{E}}^{-4/5}\left(\frac{m_1}{10^8M_\odot}\right)^{-0.1}\left(\frac{R}{R_\text{Q}}\right)^{21/20}.
\end{align}
For the $\beta$ disk, we assume that $\beta\ll 1,b=1,R\leq 10^3  R_\text{s}$, then Eq.~\eqref{Sigma} gives the density distribution
\begin{align}\label{rho}
    \rho_{\beta}&=\frac{\Omega\,\Sigma_{\beta}(R)}{2 c_\text{s}}\exp{\left(-\frac{Z^2}{2H^2_{\text{in}}}\right)}\nonumber\\
    &\approx 7.98\times10^{-10}\,\left(\frac{\alpha}{0.3}\right)^{-4/5}
    {l_\text{E}}^{-2/5}\exp{\left(-\frac{Z^2}{2H^2_{\text{in}}}\right)}\nonumber\\
    &\quad\times\left(\frac{m_1}{10^8M_\odot}\right)^{-4/5}\left(\frac{R}{R_\text{Q}}\right)^{-3/5}\text{g}\,\text{cm}^{-3},
\end{align}
the cylindrical coordinates $(R,Z)$ are related to the spherical coordinates $(r,\theta)$ via $R=r\sin{\theta}$ and $Z=r\cos{\theta}$. For the $\alpha$ disk ($\beta\ll1,b=0,R\leq 10^3  R_\text{s}$), we obtain
\begin{align}
    \rho_{\alpha}&=\frac{\Omega\,\Sigma_{\alpha}(R)}{2 c_\text{s}}\exp{\left(-\frac{Z^2}{2H^2_{\text{in}}}\right)}
    \nonumber\\
    &\approx 5.51\times10^{-11}\,\left(\frac{\alpha}{0.3}\right)^{-1}
    {l_\text{E}}^{-2}\exp{\left(-\frac{Z^2}{2H^2_{\text{in}}}\right)}\nonumber\\
    &\quad\times\left(\frac{m_1}{10^8M_\odot}\right)^{-1}\left(\frac{R}{R_\text{Q}}\right)^{3/2}\text{g}\,\text{cm}^{-3}.
\end{align}
The scale height of the inner disk can be estimated as \cite{2004ApJ...608..108G}
\begin{equation}
    H_{\text{in}}=\frac{c_\text{s}}{\Omega}\approx 8.725 \,l_\text{E}\,R_\text{s},
\end{equation}
where $c_\text{s}=\sqrt{p/\rho}$ is the isothermal sound speed and $\Omega=\sqrt{m_1/R^3}$ is the Keplerian orbital frequency.

For $R\geq  R_\text{Q}$, we adopt the disk model of \cite{2003MNRAS.339..937G}, the density is given by
\begin{equation}
\begin{aligned}
  \rho(R,Z)&=\rho(R_\text{Q},0)\left(\frac{R_\text{Q}}{R}\right)^{3}\exp\left(-\frac{Z^2}{2H^2_{\text{out}}}\right),
\end{aligned}
\end{equation}
with $H_{\text{out}}\approx \left(R/R_\text{Q}\right)^{3/2}H_{\text{in}}$ being the scale height of the outer disk.

\onecolumngrid
\section{Analytical approximation to the variations of osculating orbital elements}\label{appendix_analytical}

In this appendix, we derive analytical approximations to the changes of osculating elements during a hyperbolic encounter due to the leading PN corrections\footnote{In the vacuum GR, the fast EMRHEs (for which the dissipative effect is negligible) involving a spinning central BH should be more accurately described by the unbound geodesics in a Kerr background, which have been studied in \cite{Bini:2017pee,Hackmann:2010zz,Damgaard:2022jem,Dyson:2024qrq}.}. To the best of our knowledge, these have not been derived previously. Recall that a Keplerian hyperbolic orbit can be parametrized by six orbital elements $\{a,e,\varphi_0,\phi_0,i,t_0\}$, with
\begin{align}
r=\frac{a(e^2-1)}{1+e\cos\varphi}=a(e\cosh \xi-1),
\quad
\mathcal{M}=e \sinh \xi-\xi=\Omega (t-t_0),
\end{align}
where the true anomaly $\varphi\in[-(\pi-\arctan \sqrt{e^2-1}),\pi-\arctan \sqrt{e^2-1}]$, the eccentric anomaly $\xi\in(-\infty,\infty)$, $\mathcal{M}$ the mean anomaly, $t_0$ the periastron-crossing time, $\Omega=\sqrt{M/a^3}$, and the impact parameter $b=a\sqrt{e^2-1}$. The relation between $\varphi$ and $\xi$ is given by
\begin{align}
	\varphi=2\,\text{arctan}\left(\sqrt{\frac{1+e}{e-1}}\tanh\frac{\xi}{2}\right),
	\quad
	\cos\varphi=\frac{e-\cosh \xi}{e\cosh \xi -1},
	\quad
	\sin\varphi=\frac{\sqrt{e^2-1} \sinh \xi}{e \cosh \xi-1}.
\end{align}
The magnitude of orbital velocity is $v=a\Omega\sqrt{(1+e\cosh\xi)/(e\cosh\xi-1)}$. The conserved energy and angular momentum are $E=v^2/2-M/r=M/(2a)$ and $\mathbf{h}=\mathbf{r}\times\mathbf{v}=\sqrt{(e^2-1)a M}\,\mathbf{e}_Z$.

In the presence of additional acceleration $\mathbf{F}=F_r\mathbf{e}_r+F_\varphi \mathbf{e}_\varphi+F_Z\mathbf{e}_Z$, where $\mathbf{e}_\varphi=\mathbf{e}_Z\times\mathbf{e}_r$, the complete set of Gaussian perturbation equation for the osculating hyperbolic orbit reads
\begin{align}
	\frac{\dot{a}}{a}&=-\frac{2}{\Omega}\left(\frac{e\sin\varphi}{a\sqrt{e^{2}-1}}F_{r}+\frac{\sqrt{e^{2}-1}}{r}F_{\varphi}\right),
	\\
	\dot{e}&=\frac{\sqrt{e^2-1}}{a\Omega} \left[(\cos\varphi+\cosh\xi)F_\varphi+\sin\varphi F_r\right],
	\\
	\dot{\varphi}_{0}&=\frac{\sqrt{e^2-1}}{ae\Omega}\left[\left(1+\frac{ar}{b^2}\right)\sin\varphi F_\varphi-\cos\varphi F_r\right]-\cos i\,\dot\phi_0,
	\\
	\dot \phi_0&=\frac{r\sin (\varphi+\varphi_0+\pi/2) F_Z}{a^2\Omega\sqrt{e^2-1}\sin i},\label{orbit plane precession}
	\\
	\dot i&=\frac{r\cos (\varphi+\varphi_0+\pi/2) F_Z}{a^2\Omega\sqrt{e^2-1}}, \label{i changing rate}
	\\
	\dot{\mathcal{M}} &=\Omega+
	(1-e\cosh \xi)\frac{a \cosh \xi\,\dot e-(1-e\cosh \xi)\dot a}{ae \sinh\xi}+\sinh\xi \,\dot e
	,\\
    \dot \varphi &= \frac{a^2\Omega\sqrt{e^2-1}}{r^2}-(\dot\varphi_0+\cos i\,\dot\phi_0).
\end{align}

We consider the 1PN and 2.5PN acceleration given by Eq.~\eqref{F_1PN} and \eqref{F_2.5PN}, as well as the leading-order acceleration due to the spin angular momentum $\mathbf{J}=J\,\mathbf{e}_z$ of the primary body \cite{PhysRevD.52.821}
\begin{equation}\label{spin}
	\begin{aligned}
		\mathbf{F}^\text{(spin)}=(1+q)^2\frac{1+\sqrt{1-4\nu}}{4r^{3}}\{12\mathbf{e}_r\left[(\mathbf{e}_r\times\mathbf{v})\cdot\mathbf{J}\right]
		-(7+\sqrt{1-4\nu})\,\mathbf{v}\times\mathbf{J}
		+(9+3\sqrt{1-4\nu})\,\dot{r}\left(\mathbf{e}_r\times\mathbf{J}\right)\},
	\end{aligned}
\end{equation}
and the relative acceleration due to a quadrupolar Newtonian potential of the primary body (with its symmetry axis parallel to $\mathbf{e}_z$),
\begin{align}\label{quadrupole}
	\mathbf{F}^\text{(quad)}=-\nabla\Phi_\text{quad},
	\quad
	\Phi_\text{quad}=\left(\frac{M}{m_1}\right)\frac{Q}{r^3}\frac{1-3\cos^2\theta}2.
\end{align}
This can also be applied to the spin-induced quadrupole moment of Kerr BH (with $J=m_1^2\chi$) \cite{Poisson:1997ha} $Q=-m_1^3\chi^2$. We assume a fixed spin or symmetry axis $\mathbf{e}_z$ for the primary body and neglect the spin of the secondary body, which is a reasonable approximation if the mass ratio $q=m_2/m_1$ is sufficiently small.

\begin{figure*}[t]
	\centering
    \includegraphics[width=1.0\textwidth]{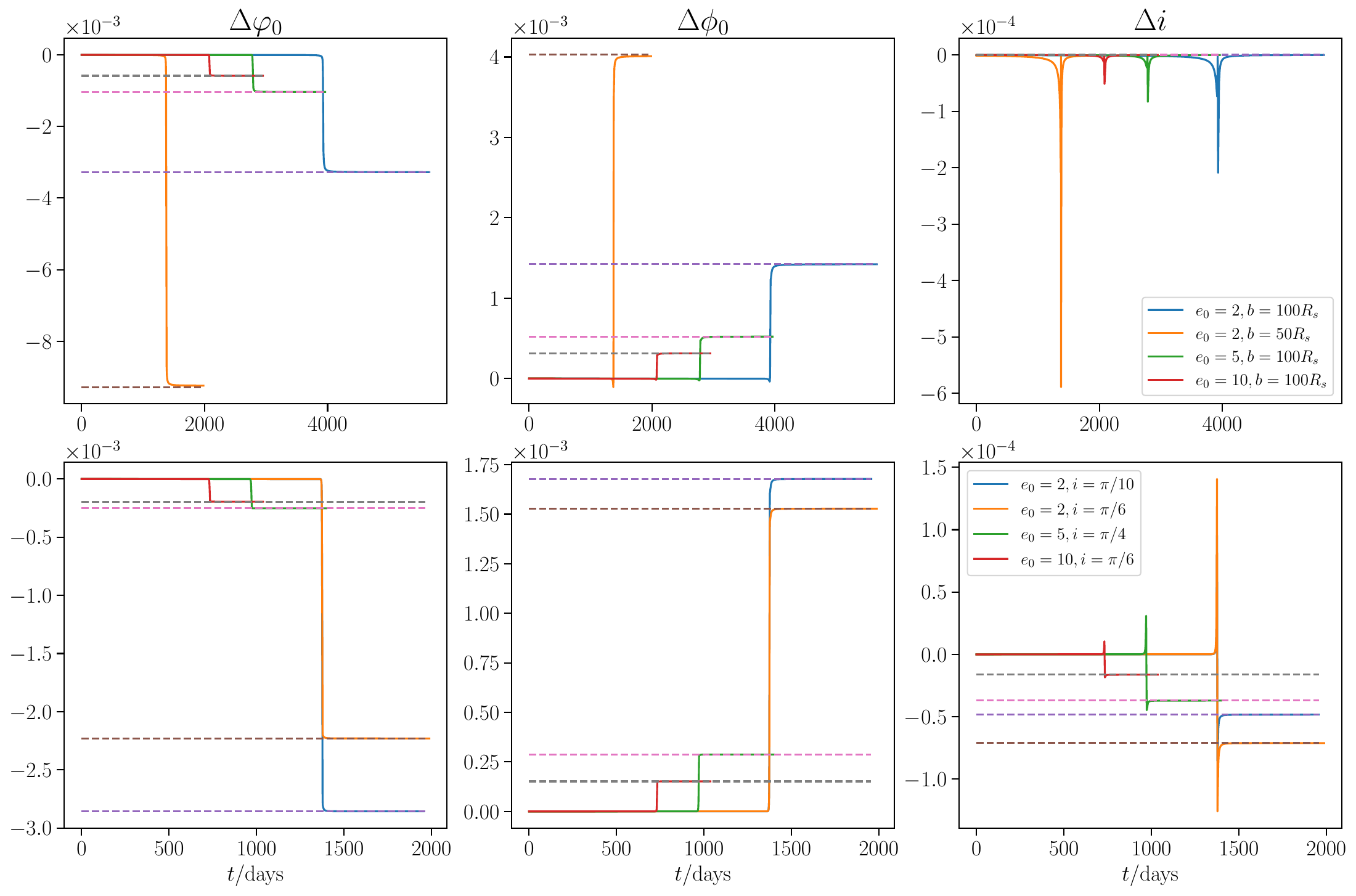}
	\caption{The variation of angular osculating elements $\Delta \mathcal{X}=\mathcal{X}(t)-\mathcal{X}(0)$ during a hyperbolic encounter under $\mathbf{F}^\text{(spin)}$ (upper row) with $m_1=10^8 \, M_{\odot}$, $m_2=10^3 \, M_{\odot}$, $\varphi_0(t=0)=\pi/6$, $J=0.6 \, {m}^2_1$, $i_0=\pi/6$ and different $(e_0,b)$, or under $\mathbf{F}^\text{(quad)}$ (lower row) with $m_1=10^8 \, M_{\odot}$, $m_2=10^3 \, M_{\odot}$, $\varphi_0(t=0)=\pi/6$, $Q=10^{-2} \,M_{\odot}\, {\text{pc}}^2$, $b=50 R_\text{s}$ and different $(e_0,i_0)$. The analytical approximations \eqref{spin_Delta_varphi0}, \eqref{spin_Delta_phi0}, \eqref{spin_Delta_i} and \eqref{quad_Delta_varphi0}, \eqref{quad_Delta_phi0}, \eqref{quad_Delta_i} to the total variations are shown as dashed curves.}
\label{fig:spin_and_quad}
\end{figure*}

The change of an orbital element $\mathcal{X}$ after the encounter is approximately given by $\Delta \mathcal{X}\approx \int_{-\infty}^\infty dt\,\dot{\mathcal{X}}$, with the osculating elements involved in $\dot{\mathcal{X}}$ fixed to their initial values. The results obtained from the Gaussian perturbation equation are
\begin{align}
			\frac{(\Delta \varphi_0)_\text{1PN}}{\frac{M}{a}}
			&=
			\frac{\sqrt{e^2-1} \left[(5 \nu +2)e^2-5 \nu +4\right]+12 e^2 \arctan \frac{e+1}{\sqrt{e^2-1}}}{e^2 \left(e^2-1\right)}\label{1PN_Delta_varphi0}
			,
			\\
			\frac{(\Delta \varphi_0)_\text{spin}}{\frac{(1+q)^2J\cos i}{e^2 \left(e^2-1\right)^2(a^3M)^{1/2}}}
			&=
			e^2 \left(2 \nu  \cos 2 \varphi_0+\nu +2 \sqrt{1-4 \nu }+2\right)
			+6e^2g(e)\sqrt{e^2-1}\left(\nu-2 \sqrt{1-4 \nu}-2\right)
			\nonumber
			\\
			&\quad+
			e^4 \left(-\nu\cos 2 \varphi_0+\nu -4 \sqrt{1-4 \nu }-4\right)-\nu  \cos 2 \varphi_0-2 \nu +2 \sqrt{1-4 \nu }+2
			\label{spin_Delta_varphi0},
			\\
			\frac{(\Delta \phi_0)_\text{spin}}{\frac{(1+q)^2J}{e^2 \left(e^2-1\right)^2(a^3M)^{1/2}}}
			&=
			e^2 \left(-2 \nu  \cos 2 \varphi_0+\nu -2 \sqrt{1-4 \nu }-2\right)
			+2e^2g(e)\sqrt{e^2-1}\left(-\nu+2+2 \sqrt{1-4 \nu}\right)
			\nonumber
			\\
			&\quad +
			e^4 \left(\nu  \cos 2 \varphi_0-\nu +2 \sqrt{1-4 \nu }+2\right)+\nu  \cos 2 \varphi_0
			\label{spin_Delta_phi0},
			\\
			\frac{(\Delta i)_\text{spin}}{\frac{(1+q)^2J}{(a^3M)^{1/2}}}
			&=
			-\frac{\nu\sin 2 \varphi_0}{e^2}\sin i 
			\label{spin_Delta_i},
			\\
			\frac{(\Delta \varphi_0)_\text{quad}}{\frac{Q}{a^2m_1}}
			&=
			\frac{2 \left(1-e^2\right)  \left[\left(2 e^2-1\right) \cos 2 i+1\right]\cos 2 \varphi_0-e^2 \left[3 \left(4 e^2+1\right) \cos 2 i+8 e^2+1\right]}{4 e^4 \left(e^2-1\right)^{3/2}}
			-g(e)\frac{15 \cos 2 i+9}{2\left(e^2-1\right)^2}
			\label{quad_Delta_varphi0},
			\\
			\frac{(\Delta \phi_0)_\text{quad}}{\frac{Q}{a^2m_1}}
			&=\frac{\sqrt{e^2-1} \left[\left(e^2-1\right) \cos 2 \varphi_0+3 e^2\right]+6 e^2 g(e)}{e^2 \left(e^2-1\right)^2}\cos i
			\label{quad_Delta_phi0},
			\\
			\frac{(\Delta i)_\text{quad}}{\frac{Q}{a^2m_1}}
			&=-\frac{\sin 2\varphi_0}{2e^2 \sqrt{e^2-1}}\sin 2i
			\label{quad_Delta_i},\quad
            (\Delta e)_\text{quad}=\frac{(e^2-1)\tan i}{e}(\Delta i)_\text{quad},
			\\
			\frac{\left(\frac{\Delta e}{e}\right)_\text{2.5PN}}{ \left(\frac{M}{a}\right)^{5/2}}
			&=
			-\frac{2\nu}{45\left(e^2-1\right)^2}
			\left[1069+72 e^2+\frac{134}{e^2}+\frac{6 \left(121 e^2+304\right)g(e) }{\sqrt{e^2-1}}\right]
			,
			\\
			\frac{\left(\frac{\Delta a}{a}\right)_\text{2.5PN}}{\left(\frac{M}{a}\right)^{5/2}}
			&=
			\frac{4\nu}{45\left(e^2-1\right)^4}
			\left[-602+673 e^4-71 e^2
			+6 \left(37 e^4+292 e^2+96\right) g(e)\sqrt{e^2-1}
			\right]
			,
		\end{align}
where $g(e)=\arctan \frac{e-1}{\sqrt{e^2-1}}+\arcsin \frac{1}{e}$. Also, $(\Delta a)_\text{1PN}=(\Delta e)_\text{1PN}=(\Delta a)_\text{spin}=(\Delta e)_\text{spin}=(\Delta a)_\text{quad}=0$. These are compared with numerical results in Fig.~\ref{fig:spin_and_quad}, showing good agreements in all cases. In principle one can add corrections of higher PN order, but a higher-order perturbative treatment of the Gaussian perturbation equation would be needed to handle these properly (similar to the case of bound orbit \cite{Mora:2003wt}). Note that a 1PN-accurate quasi-Keplerian parametrization of the unbound orbit was derived in \cite{1985AIHPA..43..107D} (which can be obtained from the parametrization of bound orbit by the analytic continuation: $\xi\to i\xi$, $a\to -a$, $\Omega\to -i\Omega$ and $\sqrt{1-e^2}\to i\sqrt{e^2-1}$), when the 1PN conserved quantities are evaluated using the full initial conditions corresponding to the ingoing Keplerian osculating orbit with elements $\{a,e\}$ and without making the $r\to \infty$ approximation, it gives a total periastron shift which is generally more accurate than Eq.~\eqref{1PN_Delta_varphi0}, but the relative difference is below $\mathcal{O}(M/a)$. The results above can be used to improve the estimation of relativistic dynamical friction \cite{PhysRevD.95.064014,Chiari:2022kas}. They can also be used to derive the leading relativistic corrections to the linear memory\footnote{In the case of bound orbit, the GW waveform only has a nonlinear memory \cite{Favata:2011qi,Gasparotto:2023fcg,2024arXiv240609228I}.} of a hyperbolic encounter, which for $q\ll1$ is given by \cite{1987Natur.327..123B}
\begin{equation}
\Delta h_{ij}^\text{TT}=\Lambda_{ij,kl}(\mathbf{n})\sum_{I=1,2}
\frac{4m_I}{D}\,\Delta\left[\frac{v^k_Iv^l_I}{\sqrt{1-v_I^2}\,(1-\mathbf{v}_I\cdot\mathbf{n})}\right]\approx 
\Lambda_{ij,kl}(\mathbf{n})\,
\frac{4\mu}{D}\,\Delta\left[\frac{v^kv^l}{\sqrt{1-v^2}\,(1-\mathbf{v}\cdot\mathbf{n})}\right].
\end{equation}
The ingoing (outgoing) velocity is fixed by the initial (final) osculating orbital elements according to Eq.~\eqref{velocity}, with $\varphi=\pm (\pi-\arctan \sqrt{e^2-1})$. For example, the Newtonian-order memory for a Keplerian orbit with $i=\varphi_0=0$ is $\Delta h_+/(\frac{\sqrt{e^2-1}}{e^2}h_0)=-2(\cos 2\theta_\text{s} +3)\,\sin 2(\phi_\text{s}-\phi_0)$, $\Delta h_\times/(\frac{\sqrt{e^2-1}}{e^2}h_0)=-8\cos \theta_\text{s}\,\cos 2(\phi_\text{s}-\phi_0)$, where $h_0=\mu M/(aD)$.

Here we also present the periastron shift for a more generically parametrized 1PN binary Lagrangian, allowing deviations from the standard vacuum GR:
\begin{equation}\label{1PN_L}
	\begin{aligned}
		L=
		&\,\frac{m_1v_1^2+m_2v_2^2}2+\frac{m_1v_1^4+m_2v_2^4}8
		+\frac{m_1m_2}rA'-\frac{m_1m_2\left(m_1+m_2\right)}{2r^2}B'
		\\
		&+\frac{m_1m_2}{2r}\left[3\left(v_1^2+v_2^2\right)C'-7\mathbf{v}_1\cdot\mathbf{v}_2\,D'\right]
		-\frac{m_1m_2}{2r}\big\{(\mathbf{v}_{1}\cdot\mathbf{e}_r)(\mathbf{v}_{2}\cdot\mathbf{e}_r)E'
	 +[(\mathbf{v}_{1}\cdot\mathbf{e}_r)^2+(\mathbf{v}_{2}\cdot\mathbf{e}_r)^2]F'\big\}
		.
	\end{aligned}
\end{equation}
This can be applied to, e.g., a binary endowed with scalar/vector charges in case that the scalar/vector fields being massless. The Einstein–Infeld–Hoffman Lagrangian corresponds to $A'=B'=C'=D'=E'=1$ and $F'=0$. In the c.m. frame of the binary, the Lagrangian \eqref{1PN_L} can be written as
\begin{equation}
\frac{L}{\mu}=G_1 v^2+G_2 v^4 +G_3 \frac{v^2}{r} + G_4\frac{\dot r^2}{r}+\frac{A}{r}+\frac{B}{r^2},
\end{equation}
with $A=A' M$, $B=-B' M^2/2$, $G_1=
1/2$, $G_2=
\left(1-3 \nu\right)/8$, $G_3/M=
3(1/2-\nu)C'+(7/2)(D')^2\nu$, and $G_4/M=
E'\nu/2+F'(\nu-1/2)$. From the equation of motion we obtain the acceleration $\mathbf{F}^\text{(0PN)}=-\frac{A}{2G_1}\frac{\mathbf{e}_r}{r^2}$, and
\begin{align}
	\frac{\mathbf{F}^\text{(1PN)}}{1/r^2}=
	\left[(2 A G_2-G_1 G_3-2 G_1 G_4)v^2
	+3 G_1 G_4(\dot r)^2+\frac{A G_3+A G_4-2 B G_1}{r}\right]\mathbf{e}_r
	+(4 A G_2+2 G_1 G_3)\,\dot r\mathbf{v}
	.
\end{align}
For an elliptical orbit, the precession angle of the periastron per orbital period $T=2\pi/\Omega$ is given by
\begin{equation}
	\frac{\varphi_0(t+T)-\varphi_0(t)}{T}=
	\frac{A [2 A G_2+G_1 (2 G_3+G_4)]+2 B G_1^2}{2\sqrt{2a^5 AG_1^5}\left(1-e^2\right)}
	.
\end{equation}
In the vacuum GR, $\frac{\varphi_0(t+T)-\varphi_0(t)}{T}=
\frac{3 M^{3/2}}{a^{5/2} \left(1-e^2\right)}
$. For a hyperbolic orbit, we obtain the change of $\varphi_0$ during the scattering:
\begin{equation}
	\begin{aligned}
		\frac{\Delta \varphi_0}{(AG_1^2a\sqrt{e^2-1})^{-1}}=&
		\left(4 A^2 G_2+4 A G_1 G_3+2 A G_1 G_4+4 B G_1^2\right)\arctan\frac{e+1}{\sqrt{e^2-1}}
		\\
		&-2 A^2 G_2+A G_1 G_3+A G_1 G_4
	 \,(4 A^2 G_2+A G_1 G_3+2 B G_1^2)\,e^{-2}.
	\end{aligned}
\end{equation}
In the vacuum GR, Eq.~\eqref{1PN_Delta_varphi0} is recovered. The results for elliptical and hyperbolic orbits match in the parabolic limit $e=1$ with $\lim_{\epsilon\to 0^+}\left\{[\varphi_0(t+T)-\varphi_0(t)]_{e=1-\epsilon}-(\Delta \varphi_0)_{e=1+\epsilon}\right\}=0$.

Finally we consider a power-law radial acceleration: $\mathbf{F}=-A\,r^{-n}\,\mathbf{e}_r$, with $A$ and $n\in\mathbb{R}$ being constants. For an elliptical orbit, we obtain
\begin{equation}
\begin{aligned}
\frac{\varphi_0(t+T)-\varphi_0(t)}{T\left(\frac{Aa^{-n}}{2}\right)\sqrt{\frac{a(1-e^2)}{M}}}&=\sum_{\lambda=\pm}\left[
\frac{\lambda-e}{e}(1-\lambda e)^{-n} \, _2F_1\left(\frac{1}{2},n;1;\frac{2 e}{e- \lambda}\right)\right.
\\
&\left.\quad +\frac{(1-\lambda e)^{-n}}{(e+\lambda) e}\left\{[4 \lambda e (n-1)+2] \, _2F_1\left(\frac{1}{2},n;2;\frac{2 e}{e-\lambda}\right)+3 \lambda(e-\lambda) \, _2F_1\left(-\frac{1}{2},n;2;\frac{2 e}{e-\lambda}\right)\right\}
\right]
,
\end{aligned}
\end{equation}
where $\,_2F_1\left(a,b;c;z\right)$ is the Gauss hypergeometric function. For a hyperbolic orbit, we obtain
\begin{equation}\label{power-law}
\begin{aligned}
\frac{\Delta \varphi_0}{A \frac{2^n a^3(a e)^{-n-1}\sqrt{e^2-1}}{n \left(n^2-1\right)M}}&=
-n(n-1)\,F_1\left(n+1;n,n;n+2;\frac{i \sqrt{e^2-1}+1}{e},\frac{e}{i \sqrt{e^2-1}+1}\right)
\\
&\quad
+2 e (n^2-1)\,F_1\left(n;n,n;n+1;\frac{i \sqrt{e^2-1}+1}{e},\frac{e}{i \sqrt{e^2-1}+1}\right)
\\
&\quad-n(n+1)\,F_1\left(n-1;n,n;n;\frac{i \sqrt{e^2-1}+1}{e},\frac{e}{i \sqrt{e^2-1}+1}\right)
,
\end{aligned}
\end{equation}
where $F_1\left(a;b_1,b_2;c;x,y\right)$ is the Appell function of the first kind, and the result converges only for $n>1$. Under $\mathbf{F}^\text{(1PN)}$ and a conservative radial acceleration (decaying to zero sufficiently fast as $r\to \infty$), the elements $\{a,e,i,\phi_0\}$ of ingoing and outgoing osculating orbits are identical, the deflection angle of the scattering is then given by $2\arccos \left(-\frac{1}{e}\right)-\pi+\Delta\varphi_0$.

\twocolumngrid

\bibliography{paper}
\end{document}